\documentclass[12pt]{article}
\usepackage[utf8]{inputenc}

\usepackage{cite}
\usepackage[colorlinks=true,linkcolor=blue, linktoc=page, urlcolor=blue,citecolor=magenta]{hyperref}
\usepackage{amsmath}
\usepackage{amssymb}

\numberwithin{equation}{section}

\usepackage{amsfonts}
\usepackage{dsfont}

\usepackage{graphicx}
\usepackage[dvipsnames]{xcolor}

\def\calc{{\cal C}}
\def\cald{{\cal D}}
\def\cale{{\cal E}}

\def\call{{\cal L}}
\def\calm{{\cal M}}

\def\calo{{\cal O}}

\def\a{\alpha}
\def\b{\beta}
\def\g{\gamma}
\def\G{\Gamma}

\def\e{\epsilon}

\def\l{\lambda}

\def\m{\mu}
\def\n{\nu}

\def\O{\Omega}

\def\r{\rho}

\def\x{\xi}

\def\Gu{{\Gamma}}

\newcommand{\os}[2]{{\overset{\,\scalebox{0.5}{($#2$)}}{#1}}{}}

\begin{document}

	\begin{titlepage}
		\vskip 1.8 cm
		
		\begin{center}{\huge \bf A 3+1 formulation of the $1/c$ expansion of General Relativity}\\
			
		\end{center}
		\vskip 1cm
		
		\begin{center}{\large {{\bf Mahmut Elbistan$^1$, Efe Hamamc\i$^1$,\\ \vspace{0.1cm}  Dieter Van den Bleeken$^{1,2}$ and Utku Zorba$^1$}}}\end{center}
		
		\vskip 0.6cm
		
		\begin{center}
			1) Physics Department, Boğaziçi University\\
			34342 Bebek / Istanbul, Turkey
			
			\vskip 0.3cm
			
			2) Secondary address:\\
			Institute for Theoretical Physics, KU Leuven\\
			3001 Leuven, Belgium
			
			\vskip 1cm
			
			\texttt{mahmut.elbistan, efe.hamamci, dieter.van, utku.zorba\, @boun.edu.tr}
		\end{center}
		\vskip 0.5cm \centerline{\bf Abstract} \vskip 0.2cm \noindent 
			Expanding General Relativity in the inverse speed of light, $1/c$, leads to a nonrelativistic gravitational theory that extends the Post-Newtonian expansion by the inclusion of additional strong gravitational potentials. This theory has a fully covariant formulation in the language of Newton-Cartan geometry but we revisit it here in a 3+1 formulation. The appropriate 3+1 formulation of General Relativity is one first described by Kol and Smolkin (KS), rather than the better known Arnowitt-Deser-Misner (ADM) formalism. As we review, the KS formulation is dual to the ADM formulation in that the role of tangent and co-tangent spaces get interchanged. In this 3+1 formulation the $1/c$ expansion can be performed in a more systematic and efficient fashion, something we use to extend the computation of the effective Lagrangian beyond what was previously achieved and to make a number of new all order observations.     \\
			
			\begin{center}
		
			\end{center}
	\end{titlepage}

\thispagestyle{empty}
\addtocounter{page}{-1}
		
		\tableofcontents

\newpage
		
\section{Introduction}\label{intro}
The geometric formulation of nonrelativistic physics has experienced a resurgence of interest in the last few years, see e.g. \cite{Bergshoeff:2022eog, Grosvenor:2021hkn, Oling:2022fft, Toappearreview} for a recent overview of the relevant geometries, symmetries and some applications. Formulating the nonrelativistic expansion of general relativity in a geometric fashion has the advantage that it keeps, by construction, general coordinate invariance manifest. Building on \cite{Dautcourt:1996pm} this idea was applied to the Post-Newtonian (PN) expansion in \cite{Tichy:2011te}, reformulating it in terms of Newton-Cartan geometry. The PN expansion, see e.g. \cite{poisson_will_2014} for an introduction, is an expansion around flat space-time and is for this reason a double expansion: nonrelativistic, i.e. an expansion in the inverse speed of light $1/c$, as well as weakly coupled, i.e. an expansion in Newton's constant $G_\mathrm{N}$. From the geometric perspective the assumption of weak coupling was implemented in \cite{Dautcourt:1996pm, Tichy:2011te} as an assumption on the nonrelativistic connection. In \cite{VandenBleeken:2017rij} this assumption was relaxed, which extends the nonrelativistic theory to so called twistless torsional Newton-Cartan geometry. Physically speaking, this extended theory adds to Newtonian gravity an additional gravitational potential that describes strong gravitational time dilation effects. Rather than an expansion around flat space-time, it is an expansion, in powers of $1/c^2$, around an arbitrary static space-time \cite{VandenBleeken:2019gqa} and as such extends the PN expansion. One of the advantages of the $1/c^2$ expansion is that an expansion of the Einstein equations goes hand in hand with an expansion of the Einstein-Hilbert Lagrangian \cite{Hansen:2019pkl, Hansen:2020pqs}. This makes it possible to formulate the expansion -- truncated up to a given order -- as a self-consistent nonrelativistic theory with a fully geometric/covariant action and associated variational principle. 

Despite the elegance of the geometrically formulated $1/c^2$ expansion \cite{VandenBleeken:2017rij, Hansen:2020pqs}, manifest space-time coordinate invariance comes with a cost, in that it introduces a number of fields which are pure gauge, and this increases the complexity of the expansion. Some work has been done to streamline this \cite{Cariglia:2018hyr, Hansen:2020wqw, Hansen:2019svu} or to organize it based on symmetry considerations \cite{Hansen:2019pkl,Bergshoeff:2019ctr, Gomis:2019sqv,Ekiz:2022wbi}, but this has so far not led to higher order results.  

In this paper we revisit the $1/c$ and $1/c^2$ expansions of general relativity in a 3+1 formulation, where an explicit choice of time coordinate is made. Although the theory loses some of its elegance in such formalism, it keeps some key features: manifest invariance under spatial coordinate transformations as well as an action with well defined variational principle. Furthermore, it has the advantage that it makes the physical degrees of freedom more explicit and the expansion more transparent. This allowed us first of all to carry out the expansion to higher order than before. It also reveals that the $1/c$ expansion is equivalently an expansion in time derivatives around a quasi-stationary relativistic background, something which was pointed out at leading order in \cite{Ergen:2020yop}. Also the relation between the $1/c$ expansion and the $1/c^2$ expansion, a consistent truncation to even orders, becomes more clear so that we could use it to devise an algorithm that constructs the $1/c^2$ expansion to power $c^{-2N}$ from the $1/c$ expansion to power $c^{-N}$. 

A 3+1 formulation of Newton-Cartan theory was recently considered in \cite{Vigneron:2020ewy, Vigneron:2020rwy}, for use in Newtonian cosmology. Rather than introducing the 3+1 formulation at the nonrelativistic level, we will start from General Relativity (GR) in 3+1 form and then expand. The most familiar 3+1 formulation of GR is that of Arnowitt, Deser and Misner (ADM) \cite{Arnowitt:1962hi}, see e.g. \cite{gourgoulhon20123+} for a detailed introduction. In ADM form the 4d relativistic metric is
\begin{equation}
ds^2=-c^2N^2 dt^2+h_{ij}(dx^i+c N^i dt)(dx^j+c N^j dt)\label{ADMnat}
\end{equation}
where the fields $(N,N^i,h_{ij})$ depend on time $t$ as well as the spatial coordinates $x^i$. Mathematically speaking $h_{ij}$ is the induced metric on a spatial hypersurface with tangent vectors $\partial_i$. The basis in tangent space is completed by the vector $u=N^{-1}(c^{-1}\partial_t-N^i\partial_i)$, which is normal to the hypersurface. The factors of the speed of light $c$ are naturally there for dimensional reasons. Now observe that the ADM decomposition becomes singular in the $c\rightarrow \infty$ limit, since $u$ is then no longer linearly independent of the $\partial_i$. This implies the ADM decomposition is not a suitable starting point for a $1/c$ expansion. (The same feature makes it\footnote{Remark that the introduction of $c$ on dimensional grounds, as in \eqref{ADMnat}, is not the unique appearance of powers of $c$ such that the frame $\{u, \partial_i\}$ remains non-degenerate in the $c\rightarrow 0$ limit. One could for example consider replacing $N^i\rightarrow c^{-1}N^i$ in \eqref{ADMnat} and the frame would remain non-degenerate as $c$ approaches 0.} a good starting point for the ultra relativistic expansion in small $c$, see e.g. \cite{Hansen:2021fxi, Campoleoni:2022ebj, Sengupta:2022jlx}). A 3+1 decomposition suitable for the $1/c$ expansion was introduced by Kol and Smolkin (KS) in \cite{Kol:2010si}. In KS form the 4d relativistic metric reads
\begin{equation}
ds^2=-M^2(cdt+C_idx^i)^2+h_{ij}dx^idx^j\label{KSintro}
\end{equation}
where again the fields $(M,C_i,h_{ij})$ depend on $t$ and $x^i$. This decomposition is based on a preferred basis for the cotangent space. The basis $dx^i$ for the spatial subspace gets extended to a basis of the total 4 dimensional cotangent space by the one-form $n=M(cdt+C_i dx^i)$. One sees that this basis remains generating in the limit $c\rightarrow \infty$ since $n$ remains linearly independent of the $dx^i$. This indicates that the KS formulation is the appropriate 3+1 decomposition for the nonrelativistic expansion. Kol and Smolkin originally introduced the decomposition \eqref{KSintro} in \cite{Kol:2007bc} in the context of the PN expansion. In that case, where one assumes $M$ to be of the weak field form $M=1-\frac{2\varphi}{c^2}+\calo(c^{-4})$, the KS ansatz, also known as the Kaluza-Klein decomposition, greatly simplifies (part of) the PN expansion, see e.g. \cite{Porto:2016pyg} for a review. 

In this work, where we consider the $1/c$ expansion and make no weak field assumption, we will allow the fields $(M, C_i, h_{ij})$ to be arbitrary analytic functions in $1/c$, which in particular means that $M=\os{M}{0}+\os{M}{1}c^{-1}+\os{M}{2}c^{-2}+\calo(c^{-4})$, with $\os{M}{k}$ arbitrary functions of $t, x^i$ (only constrained by the expanded Einstein equations). While we can relate $\os{M}{2}$ to the Newtonian potential $\varphi$, $\os{M}{0}$ and $\os{M}{1}$ are additional potentials not present in the PN expansion, capturing strong nonrelativistic gravitational effects. A similar observation holds for the expansion of $C_i$ and $h_{ij}$. The precise relation to the PN expansion (up to 1PN order) is worked out in section \ref{evexplsec}, see also table \ref{fieldstable}. 

Our paper is organized as follows. In section \ref{dualsec} we review both the ADM and KS decomposition of GR, we do so in a way that allows to treat both decompositions simultaneously and which emphasizes their dual nature. In section \ref{secsym} we focus on the KS formulation and discuss the Einstein equations as well as the diffeomorphism symmetry and the associated Noether/Bianchi identities from the point of view of this 3+1 decomposition. At the end of the section, in subsections \ref{Poissec} and \ref{expltimesec} we motivate and perform the convenient field redefinition $M=e^{\frac{\psi}{2}}$, $h_{ij}=e^{-\psi}\gamma_{ij}$ and make factors of $c$ and time derivatives explicit. In section \ref{expsec} we then come to the $1/c$ expansion. The new formalism allows us to make a number of new observations: we show how all subleading equations take the form of a sourced linear second order PDE and compute the linear differential operator that is the same at all orders. We also discuss how by a convenient choice of gauge all the subleading coefficients in the expansion of $\gamma_{ij}$ can be made trace-less. We then compute the Lagrangian of the expansion up to NNLO, i.e. up to order $c^{-2}$. Previously only the leading order, i.e. order $c^{0}$, had been computed \cite{Ergen:2020yop}. In section \ref{evsec} we recall how the $1/c^2$ expansion is a consistent truncation of the $1/c$ expansion to even orders. Here the use of the 3+1 formulation makes the connection to the $1/c$ expansion more transparent and allows us to derive an algorithm, that we call the {\it shuffling algorithm}, with which the $1/c^2$ expansion up to order $c^{-2N}$ can be constructed from the $1/c$ expansion up to order $c^{-N}$. At NLO, i.e. order $c^{-2}$, we recover the results of \cite{VandenBleeken:2017rij, Hansen:2020pqs} and spell out the precise relation. The shuffling algorithm then allows us to compute the Lagrangian up to NNLO, i.e. order $c^{-4}$, for the first time. This is the order where the 1PN correction to Newtonian gravity finds itself and we show how indeed, upon setting to zero a number of additional potentials, the $1/c^2$ expansion reproduces the well-known 1PN equations. We conclude in section \ref{discsec} where we also point towards possible future directions. Finally there are three appendices. In appendix \ref{techap} we spell out two technical derivations of results used in the main text, in appendix \ref{traceapp} we provide some extra terms present outside the traceless gauge and the (un-expanded) equations of motion with powers of $c$ and time derivatives made explicit are written down in appendix \ref{eomapp}.

\section{Two dual 3+1 formulations of GR}\label{dualsec}
Let $\calm$ be a $d+1$ dimensional manifold equipped with a Lorentzian metric $g_{\m\n}$. We'll denote the associated Levi-Civita connection with $\nabla$.
\subsection{Choice of coordinates}
Intuitively a (local) $d+1$ formulation would amount to a (local) choice of coordinates $(x^\mu)=(t,x^i)$ with $t$ a 'time coordinate' and the $x^i$ 'spatial coordinates'. We could try to make this more precise by requiring $\partial_t$ to be a time-like vector and $\partial_i$ to be space-like vectors. Although imposing both these conditions is always possible, e.g. in Gaussian normal (GN) coordinates, this would restrict the allowed set of coordinates very much. A weaker choice, which we'll call Arnowitt-Deser-Misner (ADM) coordinates \cite{Arnowitt:1962hi}, only requires the $\partial_i$ to be spatial. Let us however point out that this automatically implies that the one-form $dt$ will be time-like. So in this case we see that it is a mix of the coordinate basis of tangent vectors and that of co-tangent vectors which is required to be space/time-like respectively. This then suggests a naturally dual alternative, which is to choose coordinates such that the $dx^i$ are space-like which then implies that $\partial_t$ is time-like. We'll call such coordinates Kol-Smolkin (KS) coordinates \cite{Kol:2010si}. For a summary of the definitions see table \ref{coords}. 

\begin{table}\begin{center}
	\begin{tabular}{c|c|c}
		& space-like & time-like\\
		\hline
		&&\\
		GN & $\partial_i\,,\ dx^i$ & $\partial_t\,,\ dt$\\
		&&\\
		ADM& $\partial_i$ & $dt$\\
		&&\\
		KS& $dx^i$ & $\partial_t$
	\end{tabular}\label{coords}\caption{We'll refer to a coordinate system $(t,x^i)$ as a space-time split or $d+1$ decomposition if it falls into either the ADM or KS class (the GN class is the intersection of both). They are defined by the properties of the coordinate bases of tangent and co-tangent spaces, as listed in this table.}
\end{center}
\end{table}

\paragraph{Elementary example} Before discussing these classes in full generality in the next subsection, let us illustrate them in a simple example. Consider the Minkowski metric in inertial coordinates
\begin{equation}
ds^2=-dt^2+dx^i dx^i \, .
\end{equation}
One easily verifies that $(t,x^i)$ satisfy our definition of GN coordinates. 

If we define new coordinates $(\hat t, \hat x^i)$, with $\hat t=t$ and $\hat x^i=x^i-v^i t$ for some constants $v^i$, then the metric takes the form
\begin{equation}
ds^2=-d\hat t^2+(d\hat x^i+v^id\hat t)(d\hat x^i+v^id\hat t) \, .
\end{equation}
While $\hat\partial_i$ remains spatial, $\hat \partial_t$ is no longer time-like when $v^iv^i\geq 1$. But since $\hat g^{tt}=-1$ independent of $v^i$, it follows that $d\hat t$ remains time-like. The coordinates $(\hat t,\hat x^i)$ are thus an example of ADM coordinates.
 
Alternatively we can work in coordinates $(\tilde t, \tilde x^i)$, with $\tilde t= t-w_i x^i$ and $\tilde x^i=x^i$ for some constants $w_i$. In these coordinates the metric reads
\begin{equation}
ds^2=-(d\tilde t+w_i d\tilde x^i)^2+d\tilde x^i d\tilde x^i \, .
\end{equation}
One verifies that $\tilde \partial_t$ is always time-like, while $\tilde \partial_i$ is not space-like when $w_i^2\geq 1$. But the $d\tilde x^i$ are always space-like and so $(\tilde t, \tilde x^i)$ are an example of KS coordinates for Minkowski space, independent of the values of $w_i$.

Note that indeed the ADM and KS classes are not mutually exclusive, their intersection is exactly the GN class. 

\subsection{Decomposition of the metric}
Although our aim is to decompose the space-time metric $g_{\m\n}$ in either ADM or KS coordinates, it will be interesting to introduce a formalism that can treat both at the same time. This will then also allow us to decompose the Einstein-Hilbert action simultaneously for both cases in section \ref{EHsection}. The decomposition of the metric we'll perform here will also shed light on the definitions of the previous subsection, which might have appeared a bit ad-hoc there.

We start by choosing\footnote{Mathematically speaking a Lorentzian metric introduces an O(1,\,$d$)-structure on $\calm$. The addition of the vector field $u^\m$ refines this to an O($d$)-structure, also known as an Aristotelian structure, which is a combination of compatible Galilean and Carrollian structures, see e.g. \cite{Figueroa-OFarrill:2020gpr}. The tensors defining this Aristotelian structure are $u^\m, n_\m=g_{\m\n}u^\n$, $h^{\m\n}=g^{\m\n}+u^\m u^\n$ and $h_{\m\n}=g_{\m\n}+n_\m n_\n$. } a time-like vector field $u^\mu$, which without loss of generality we can assume to be normalized: $g_{\m\n} u^\m u^\n=-1$. Of course this is equivalent to the choice of a time-like one-form $n_\m=g_{\m\n}u^\n$, which is again normalized. 

We can complete the time-like vector into a tangent frame $(u,e_i)$, with each of the $e_i$ orthogonal to $u$: $g_{\m\n}u^\m e_i^\n=0$. The relations $e^i_\m e_j^\m =\delta_j^i$, $u^\m e_\m^i=0$ then provide a dual frame $(n,e^i)$. In summary this amounts to
\begin{equation}
-u^\mu n_\n+e^\mu_i e_\nu^i=\delta^\m_\n , \qquad e_i^\mu e_\mu^j=\delta_i^j\,.\label{invrels}
\end{equation}

Note that we do not require the frame $e_i$ to be orthonormal, instead we define
\begin{equation}
h_{ij}=g_{\m\n}e^\m_i e^\n_j \, .
\end{equation}
It follows that $h^{ij}=g^{\m\n}e_\m^i e_\n^j$ is the inverse of $h_{ij}$. 

In terms of this frame and co-frame the metric and its inverse decompose as  
\begin{equation}
ds^2=g_{\m\n}dx^\m dx^\n=-n^2+h_{ij}e^ie^j \, ,\qquad g^{\m\n}\partial_\m\partial_\n=-u^2+h^{ij}e_i e_j \, .\label{split}
\end{equation}

As we will now point out, there is a natural choice of frames of the form above associated to both ADM and KS coordinates.

\subsubsection{ADM decomposition}
In ADM coordinates, where $\partial_i$ is spatial and $dt$ time-like, a natural choice of frames satisfying \eqref{invrels} is
\begin{eqnarray}
\label{basexp1}
e_i=\partial_i\, , &\quad& n=-Ndt\,,\\
\label{basexp1a}
e^i=dx^i+N^idt\,,&\quad& u=N^{-1}(\partial_t-N^i\partial_i)\, .
\end{eqnarray}
Then, the metric (\ref{split}) takes the well-known ADM form
\begin{equation} 
\label{metcase1}
ds^2 = -N^2 dt^2 +h_{ij} (dx^i + N^i dt) (dx^j + N^j dt)\,.
\end{equation}
Note that the inverse metric takes the form
\begin{equation}
g^{\m\n}\partial_\m\partial_\n=-N^{-2}(\partial_t-N^i\partial_i)^2+h^{ij}\partial_i\partial_j\,.\label{invmetcase1}
\end{equation}

\subsubsection{KS decomposition}
In KS coordinates it is $dx^i$ which is spatial while $\partial_t$ is time-like, so in this case the natural choice of frames satisfying \eqref{invrels} is
\begin{eqnarray}
e^i=dx^i \, , &\quad& u=M^{-1}\partial_t \, ,\label{basexp2}\\
\label{basexp2a}
e_i=\partial_i-C_i\partial_t \, , & \quad & n=-M(dt+C_i dx^i) \, .
\end{eqnarray}
So the metric (\ref{split}) takes the following form in KS coordinates:
\begin{equation}
\label{metcase2}
ds^2 = -M^2 (dt + C_i dx^i)^2 + h_{ij} dx^i dx^j \, ,
\end{equation}
while the inverse metric is
\begin{equation}
g^{\m\n}\partial_\m\partial_\n=-M^{-2}\partial_t^2+h^{ij}(\partial_i-C_i\partial_t)(\partial_j-C_j\partial_t)\, .\label{invmetcase2}
\end{equation}
Comparing the forms (\ref{metcase1}, \ref{invmetcase1}) with (\ref{metcase2}, \ref{invmetcase2}) one explicitly sees how they are dual, in the sense that the role of metric and inverse metric -- i.e. tangent space and co-tangent space --  get interchanged.  

We should emphasize that the spatial metric $h_{ij}$ in the KS form of the metric \eqref{metcase2} is different from the spatial metric $h_{ij}$ in the ADM form of the metric \eqref{metcase1}. Indeed, in case the coordinates $(t,x^i)$ are in the GN class, i.e. of both ADM and KS type, they are related as
\begin{equation}
h_{ij}^{\mathrm{ADM}}=h_{ij}^{\mathrm{KS}}-M^2C_iC_j\quad\mbox{and}\quad h^{ij}_\mathrm{KS}=h^{ij}_{\mathrm{ADM}}-N^2 N^iN^j \, .
\end{equation}
It is $h_{ij}^{\mathrm{ADM}}$ which has the geometrical interpretation of the pull-back of the Lorentzian metric $g_{\m\n}$ on the constant $t$ hyper-surfaces. Although $h_{ij}^{\mathrm{KS}}$ does not have this natural geometric interpretation it is, by construction, a well-defined Riemannian metric on the constant $t$ hyper-surfaces when $(t,x^i)$ are in the KS class. Only if $(t,x^i)$ are in the ADM class will the constant $t$ hypersurfaces be spatial so that then $h_{ij}^{\mathrm{ADM}}$ is Riemannian.

\subsection{Decomposition of the connection}
Given a choice of frame and dual frame that satisfy \eqref{invrels} one has the following identities
\begin{align}
n_\m\nabla_\n u^\m&=u^\m\nabla_\n n_\m=0 \, , \label{id1}\\
e^\m_i\nabla_\n e_\m^j&=-e_\m^j\nabla_\n e^\m_i \, ,\label{id2}\\
e^\m_i\nabla_\n n_\m&=-n_\m\nabla_\n e^\m_i\, , \\
e^i_\m\nabla_\n u^\m&=-u^\m\nabla_\n e^i_\m\, .\label{id4}
\end{align}
Furthermore in either the ADM or KS case the natural frames (\ref{basexp1}, \ref{basexp1a}) or (\ref{basexp2}, \ref{basexp2a}) satisfy the further identities
\begin{equation}
e_\m^i\partial_\n e^\m_j=- e^\m_j\partial_\n e_\m^i=0,  \label{vielvan}
\end{equation}
Our results below hold for more general frames than  (\ref{basexp1}, \ref{basexp1a}) or (\ref{basexp2}, \ref{basexp2a}), as long as we assume them to satisfy the additional condition \eqref{vielvan} in addition to \eqref{invrels}.

The identities listed above are then sufficient to decompose the Levi-Civita connection of $g_{\m\n}$ in the independent components\footnote{Note that 
	when combined with \eqref{id2} the assumption \eqref{vielvan} results in $	-e^\m_i\nabla_\n e_\m^j=e_\m^j\nabla_\n e^\m_i=e_\m^j\Gamma_{\r \n}^\m e^\r_i$. This in turn guarantees that $\hat\Gamma_{ij}^k$ is symmetric in $ij$ since it can be written as $	\hat\Gamma_{ij}^k= e_i^\m e_j^\n e_\r^k\Gamma_{\m\n}^\r$.}
\begin{align}
K_{ij}&=e_i^\m e_j^\n \nabla_{\m}n_\n \, ,\label{Kdef}\\
\Omega_i&=u^\m e_i^\n\nabla_\m n_\n \, ,\label{kadef}\\
\Delta_i^j&=u^\m e^j_\n \nabla_\m e^\n_i-K_i{}^j \, ,\label{Delta1def}\\
\hat\Gamma_{ij}^k&=e_\m^k e_i^\n \nabla_\n e_j^\m \,.\label{Gbardef}
\end{align}

The key point of introducing these objects is that, via (\ref{id1}-\ref{id4}), they are sufficient to decompose the covariant derivatives of the frames. In turn that allows to decompose covariant derivatives of tensors as
\begin{align}
	\nabla_\m(e^\n_i V^i)&=-n_\m u^\n \Omega_i V^i-n_\m e^\n_i(\hat{D}_tV^i+V^jK_j{}^i)+e_\m^i u^\n K_{ij}V^j+e_\m^ie^\n_j  \hat D_i V^j\label{der1}\,,\\
	\nabla_\m(e_\n^i V_i)&=-n_\m n_\n \Omega^i V_i-n_\m e_\n^i(\hat{D}_t V_i-K_{i}{}^jV_j)+e_\m^i n_\n K_{i}{}^jV_j+e_\m^ie_\n^j \hat D_iV_j\label{der2} \, .
\end{align}
It is a bit tedious but rather obvious to see how this generalizes to tensors with any number of upper and lower 'spatial' indices, e.g. $\nabla_\m(e_\n^i e^\r_j V_i{}^j)$, so we will not write out these expressions explicitly.

In (\ref{der1}, \ref{der2}) we introduced a modified covariant derivative $\hat{D}_t$ along $t$ and a modified covariant derivative  $\hat D_i$ along the $x^i$. Their precise definitions on a 'spatial' tensor $T^{i_1\ldots i_m}_{j_1\ldots j_n}(t,x)$ are
\begin{align}
\hat{D}_t T^{j_1 \ldots j_m }_{k_1 \ldots k_n} &= \hat\partial_t T^{j_1 \ldots j_m }_{k_1 \ldots k_n}  + \Delta^{j_1}_i T^{ij_2 \ldots j_m }_{k_1 \ldots k_n} + \ldots  - \Delta_{k_1}^i  T^{j_1 \ldots j_m }_{i k_2\ldots k_n} -  \ldots  \, , \label{derivegen1}\\
\hat D_i T^{j_1 \ldots j_m }_{k_1 \ldots k_n} &= \hat\partial_i T^{j_1 \ldots j_m }_{k_1 \ldots k_n} + \hat{\Gamma}_{i l}^{j_1} T^{lj_2 \ldots j_m }_{k_1 \ldots k_n} + \ldots - \hat{\Gamma}_{i k_1}^{l} T^{j_1 \ldots j_m }_{l k_2\ldots k_n}-\ldots  \, .\label{derivegen2}
\end{align}
where
\begin{equation}
\hat \partial_t=u^\mu \partial_\mu\qquad\mbox{and}\qquad \hat \partial_i=e_i^\mu\partial_\m \, .
\end{equation}
Note that $\hat D_i$ differs from the standard covariant derivative $D_i$ with respect to the Levi-Civita connection of $h_{ij}$, i.e.
\begin{align}
D_i T^{j_1 \ldots j_m }_{k_1 \ldots k_n}&= \partial_i T^{j_1 \ldots j_m }_{k_1 \ldots k_n}  + \Gamma^{j_1}_{i l} T^{lj_2 \ldots j_m }_{k_1 \ldots k_n} + \ldots  - \Gamma^l_{ik_1 } T^{j_1 \ldots j_m }_{l k_2\ldots k_n} - \ldots  \, ,\label{derivegenLC}\\
\Gamma_{ij}^k&=\frac{1}{2}h^{kl}\left(\partial_i h_{jl}+\partial_j h_{il}-\partial_l h_{ij}\right) . \label{LCdef}
\end{align}
The difference is two-fold, firstly in $\hat D_i$ the standard partial derivative $\partial_i$ is replaced by $e_i^\mu \partial_\mu$ and secondly the 'spatial connection' $\hat \Gamma_{ij}^k$ is used instead of $\Gamma_{ij}^k$. Both these differences vanish in the ADM case, but not in the KS case -- as we'll see below. It is important to point out that the modified covariant derivative \eqref{derivegen2} remains compatible with $h_{ij}$:
\begin{equation}
\hat D_i h_{jk}=0
\end{equation}
This is not so for the time-like covariant derivative, but one finds the elegant expression
\begin{equation}
\hat D_th_{ij}=2K_{(ij)} \, . 
\end{equation} 
Let us remark that similarly to the decomposition \eqref{der1} one computes that 
\begin{equation}
\nabla_\m (V u^\m+V^i e_i^\m)=\hat{D}_tV+K V+\hat D_i V^i+\Omega_i V^i \, ,\label{totder}
\end{equation}
and finally we also point out the relations
\begin{align}
\hat{\partial}_t\hat{\partial}_i-\hat{\partial}_i\hat{\partial}_t&=\Omega_i\hat\partial_t+\Delta_i^j\hat{\partial}_j\, ,\\
\hat{\partial}_i\hat{\partial}_j-\hat{\partial}_j\hat{\partial}_i&=K_{[ij]}\hat{\partial}_t \,.
\end{align}

\subsubsection{The ADM case}
When choosing the frames to be of the ADM type (\ref{basexp1}, \ref{basexp1a}) the hatted partial derivatives take the form
\begin{equation}
\hat \partial_t=N^{-1}(\partial_t-N_i\partial_i)\,,\qquad\hat\partial_i=\partial_i \, .\label{hpADM}
\end{equation}
The hatted time derivative, which in this case is hypersurface orthogonal, includes the well known shift by the vector field $N^i$ (see e.g. \cite{gourgoulhon20123+}), while the hatted spatial derivatives are simply partial derivatives.

The connection components (\ref{Kdef}-\ref{Gbardef}) become in this case
\begin{align}
K_{ij}&=N^{-1}\left(\frac{1}{2}\partial_t h_{ij}-D_{(i}N_{j)}\right) \, ,\label{Kexplicit}\\
\Omega_i&=N^{-1}\partial_i N \, ,\label{OmADM}\\
\Delta_i^j&=N^{-1}\partial_i N^j\, ,\label{DADM}\\
\hat\Gamma_{ij}^k&=\Gamma_{ij}^k \, .
\end{align}
First one recognizes in $K_{ij}$ the extrinsic curvature of the induced metric $h_{ij}$, while $\Omega_i$ is the so called Eulerian acceleration vector (see e.g. \cite{gourgoulhon20123+}). Secondly one sees that the connection $\hat \Gamma_{ij}^k$ in this case equals the Levi-Civita connection \eqref{LCdef}, so that $\hat D_i=D_i$.
Finally the object $\Delta_i^j$ plays a role as a 'connection' in the time-like covariant derivative \eqref{derivegen1}. Writing that out explicitly one gets the expression
\begin{equation}
\hat{D}_t T^{j_1 \ldots j_m }_{k_1 \ldots k_n}=N^{-1}(\partial_t-L_N) T^{j_1 \ldots j_m }_{k_1 \ldots k_n} \, ,
\end{equation}
where $L_N$ is the Lie derivative with respect to the vector field $N^i$. Remark that this also allows one to rewrite \eqref{Kexplicit} as
\begin{equation}
K_{ij}=\frac{1}{2}\hat D_t h_{ij}\, .\label{KDt}
\end{equation}

\subsubsection{The KS case}
Now we assume the frames to take the KS form (\ref{basexp2}, \ref{basexp2a}). The hatted partial derivatives now are
\begin{equation}
\hat\partial_t=M^{-1}\partial_t \, ,\qquad \hat\partial_i=\partial_i-C_i\partial_t \, .\label{hpKS}
\end{equation}
The situation is exactly opposite -- or dual -- to the ADM case \eqref{hpADM}, in that the hatted time derivative is simply (up to a prefactor) the partial derivative, while it is the spatial hatted derivatives that get a shift, this time by the one-form $C_i$.

When writing out the objects (\ref{Kdef}-\ref{Gbardef}) one finds they are naturally expressed in terms of these hatted partial derivatives\footnote{This is also true in the ADM case, as there $\partial_i=\hat{\partial}_i$ and \eqref{Kexplicit} can be re-expressed as \eqref{KDt}.}:
\begin{align}
K_{ij}&= \frac{1}{2}\hat\partial_t h_{ij}-M\hat{\partial}_{[i}C_{j]} \, ,\label{KKS}\\
\Omega_i&= M^{-1}\hat\partial_i M-M \hat \partial_tC_i \, ,\label{OmKS}\\
\Delta_i^j&= 0\, ,\label{DKS}\\
\hat\Gamma_{ij}^k&=\frac{1}{2}h^{kl}\left(\hat\partial_i h_{jl}+\hat\partial_j h_{il}-\hat\partial_l h_{ij}\right) \, . \label{GhatKS}
\end{align}
It might appear as if the introduction of the hatted partial derivatives is simply a notation to simplify the expressions (\ref{KKS}-\ref{GhatKS}). But, as we will discuss in section \ref{secsym}, these hatted derivatives are the natural ones invariant under redefinitions of the time coordinate, which is why all invariant quantities involving derivatives can be expressed in terms of them.

Let us now come to the expressions (\ref{KKS}-\ref{GhatKS}) themselves. While the symmetric part of $K_{ij}$ is analogous to the ADM expression \eqref{KDt} (see also \eqref{ptDt}), it no longer has an interpretation as extrinsic curvature. More importantly, $K_{ij}$ now also has a non-vanishing anti-symmetric part (absent in the ADM case), which takes the form of a (generalized) curvature for the 1-form potential $C_i$. The vector $\Omega_i$ can also in this KS case be interpreted as an 'acceleration vector', in that it provides a relativistic version of Newton's gravitational field vector (see sections \ref{Poissec}). Since $\Delta^i_j$ vanishes we can conclude that
\begin{equation}
\hat D_t=\hat \partial_t  \, .\label{ptDt}
\end{equation}
On the contrary, $\hat\Gamma_{ij}^k$ is in this case not the Levi-Civita connection, rather
\begin{equation}
\hat\Gamma_{ij}^k=\G_{ij}^k+\Delta_{ij}^k,\qquad \Delta_{ij}^k=-\frac{1}{2}h^{kl}\left(C_i\partial_t h_{jl}+C_j\partial_t h_{il}-C_l\partial_t h_{ij}\right)\, .\label{KSdiff}
\end{equation}
As mentioned before, this implies that in the KS case, the spatial covariant derivative \eqref{derivegen2} is not the standard one, in that $\hat{\partial}_i$ has a shift proportional to $\partial_t$, see \eqref{hpKS}, and the connection coefficients have the extra contribution $\Delta_{ij}^k$. When discussing general relativity in KS form the expression \eqref{derivegen2}, together with \eqref{hpKS} and \eqref{GhatKS} is most convenient to work with.

\subsection{Decomposition of the Einstein-Hilbert action}\label{EHsection}
To rewrite the Einstein-Hilbert action in $d$+1 form one needs to decompose the Ricci scalar. This can be done by first expressing the Ricci tensor as $R_{\m\n}
=-2(\nabla_{[\r}\nabla_{\n]} u^\r)n_\m+2(\nabla_{[\r}\nabla_{\n]} e^\r_i)e^i_\m$ and re-expressing the covariant derivatives in terms of (\ref{Kdef}-\ref{Gbardef}). Upon contraction with $g^{\m\n}$ one finds\footnote{Note that there a priory is an extra term proportional to $h^{ij}K_{[kj]}\Delta_i^{k}$, but this vanishes in both the ADM and KS case, see (\ref{OmADM}, \ref{DADM}) and (\ref{OmKS}, \ref{DKS}) respectively.}
\begin{equation}
R[g]=\hat R+ K_{ij} K^{ij}-K^2+ 2\left(\hat{D}_t K+K^2 -\hat D_i\Omega^i-\Omega_i\Omega^i\right) \, , \label{RicS}
\end{equation}
where
\begin{equation}
K=h^{ij}K_{ij}\,,\quad\hat {R}=h^{ij}\hat R_{ij}\quad\mbox{and}\quad \hat{R}_{ij}=\hat\partial_{k}\hat\G_{ij}^k-\hat\partial_{j}\hat\G_{ik}^k+\hat\G_{lk}^l\hat\G_{ij}^k-\hat\G_{ik}^l\hat\G_{lj}^k\, .
\end{equation}
Via \eqref{totder} one recognizes the term in between brackets in \eqref{RicS} as a total derivative so that\footnote{We assume boundary conditions so that the total derivative term can be ignored. Analyzing that more thoroughly could be interesting.}
\begin{equation}
S_\mathrm{EH}=\int R[g]\,\sqrt{-g}\,d^{d+1}x=\int (\hat R+ K_{ij} K^{ij}-K^2)\,n_t dt\, \sqrt{h}\,d^dx \, .\label{EH}
\end{equation}
It is gratifying to see that the Einstein-Hilbert action takes a simple and universal form for both ADM and KS type decompositions.

\subsubsection{The ADM Lagrangian}
The ADM Lagrangian \cite{Arnowitt:1962hi} is well-known:
\begin{equation}
\call_\mathrm{ADM}=N\sqrt{h}\left(R+ K_{ij} K^{ij}-K^2\right) \, . \label{ADMlag}
\end{equation}
Indeed it equals \eqref{EH} when one takes the frames of the ADM form (\ref{basexp1}, \ref{basexp1a}), since then $n_t=N$ and -- as discussed in the previous subsection --  $\hat\partial_i=\partial_i, \hat\Gamma_{ij}^k=\G_{ij}^k$, so that $\hat R=R$, the Ricci scalar of the Levi-Civita connection of $h_{ij}$. Furthermore $K_{ij}$ has the standard ADM form \eqref{Kexplicit}.

\subsubsection{The KS Lagrangian}
Since for the KS frames (\ref{basexp2}, \ref{basexp2a}) $n_t=M$ we recover from \eqref{EH} the KS Lagrangian \cite{Kol:2010si}:
\begin{equation}
\call_{\mathrm{KS}}=M\sqrt{h}\left(\hat R+ K_{ij} K^{ij}-K^2\right) \, .\label{KSlag}
\end{equation}
Although this expression might appear almost identical to the ADM Lagrangian \eqref{ADMlag}, it is rather different. This is since now, via \eqref{hpKS} and \eqref{KSdiff}, $\hat R$ has a number of additional contributions apart from $R$, the Ricci tensor of the Levi-Civita connection of $h_{ij}$. Secondly the tensor $K_{ij}$ is also different from its ADM analog, see \eqref{KKS} versus \eqref{Kexplicit}, in particular it contains a non-trivial anti-symmetric part. When we write out the time derivatives explicitly in section \ref{expltimesec}, the Lagrangian \eqref{KSlag} becomes (\ref{totlag}, \ref{L012}) and one sees some of the complexity that is elegantly packaged in the form \eqref{KSlag}.

\section{Dynamics and symmetries in the KS formulation}\label{secsym}
General relativity in the ADM formulation has been widely discussed. This is much less so for the KS formulation, so in this section we shortly discuss the Einstein equations in the KS formulation, as well as the diffeomorphism invariance of the theory. This means that in this section all definitions are those of the KS subsections of the previous section, i.e. the key definitions used here are (\ref{basexp2}-\ref{metcase2}), (\ref{hpKS}-\ref{GhatKS}) and \eqref{KSlag}. 

\subsection{Einstein equations}
The Einstein equations in KS form can be derived in two parallel and equivalent ways. The first would be to decompose the $d+1$-covariant Einstein equations using the frames $(\ref{basexp2}, \ref{basexp2a})$, the second to vary the KS Lagrangian with respect to the fields $h_{ij}, C_i, M$. Let us make the link between these two approaches explicit. The discussion in section \ref{EHsection} showed that
\begin{equation}
\call_{\mathrm{EH}}=\call_{\mathrm{KS}}+\partial_\m \Theta^\m \, .
\end{equation}
Via the KS decomposition of the metric \eqref{metcase2} and the definition of the KS frames $(\ref{basexp2}, \ref{basexp2a})$ it follows that
\begin{equation}
\delta g_{\m\n}=-2M^{-1}\delta M n_\m n_\n+2M\delta C_i n_{(\m}e_{\n)}^i+e_\m^i e_\n^j \delta h_{ij}\, .\label{dgdecomp}
\end{equation}
This implies that if we define
\begin{equation}
\delta \call_{\mathrm{EH}}=\delta\call_{\mathrm{KS}}+\partial_\m\delta \Theta^\m=\sqrt{h}M\left(2G^0 M^{-1}\delta M-2G^i M\delta C_i-G^{ij}\delta h_{ij}\right)+\partial_\m \theta^\m\, , \label{vargrav}
\end{equation}
one gets the relations
\begin{equation}
G^0=G^{\m\n}n_\m n_\n \, ,\qquad G^i=G^{\m\n}n_{(\m}e_{\n)}^i \, ,\qquad G^{ij}= G^{\m\n}e_\m^i e_\n^j\,.\label{EinstDecomp}
\end{equation}
Here $G^{\m\n}=-\frac{1}{\sqrt{-g}}\frac{\delta \call_{\mathrm{EH}}}{\delta g_{\m\n}}$ is the Einstein tensor of the metric $g_{\m\n}$. 
In summary we see that indeed there are two ways to compute the $G$'s: either via \eqref{EinstDecomp}, or via \eqref{vargrav} by a variation of $\call_{\mathrm{KS}}$. One checks that both calculations match, with the result
\begin{align}
G^0=&\,\frac{1}{2}\left(\hat R+3 K_{[ij]}K^{ij}-K_{(ij)}K^{ij}+K^2\right)\, ,\label{G0}\\
G^i=&\,\hat D_j(K^{ij}-h^{ij}K)+2K^{[ij]}\Omega_j \, ,\\
G^{ij}=&\,\hat G^{ij}-2K^{k(i}K^{j)}{}_k +K K^{(ij)}+h^{ik}h^{jl}\hat{\partial}_tK_{(kl)}-\hat D^{(i}\Omega^{j)}-\Omega^i\Omega^j \nonumber\\
&+\frac{1}{2}h^{ij}\left(3 K_{(kl)}K^{kl}-K_{[ij]}K^{ij}-K^2-2h^{kl}\hat{\partial}_t K_{kl}+2\hat{D}_i\Omega^i+2\Omega_i\Omega^i\right) \, .\label{Gij}
\end{align}
In the above $\hat G^{ij}=\hat R^{ij}-\frac{1}{2}h^{ij}\hat R$. The vacuum Einstein equations are thus equivalent to putting (\ref{G0}-\ref{Gij}) to zero. 
It is interesting to point out that $M$ appears algebraically in $\call_\mathrm{KS}$ (similarly to $N$ in the ADM formulation), so that $G^0=0$ has the interpretation of a constraint. Another consequence is that the equations do not contain a second time derivative of $M$ (while the second time derivative of both $h_{ij}$ and $C_i$ does appear). One obtains an equation for the second {\it spatial} derivatives of $M$, by taking the trace of $G^{ij}$, we will discuss that equation a bit more in section \ref{Poissec}.

One can couple matter to the theory by introducing a matter Lagrangian $\call_{\mathrm{mat}}$. This then allows to define the KS energy-momentum tensors $T$:
\begin{equation}
\delta \call_{\mathrm{mat}}=\sqrt{h}M\left(-2T^0 M^{-1}\delta M+2T^i M\delta C_i+T^{ij}\delta h_{ij}\right)\, .\label{varmatter}
\end{equation}
The Einstein equations then take the form
\begin{equation}
G^0=T^0\, ,\qquad G^i=T^i \, ,\qquad G^{ij}=T^{ij}\,.\label{Eeqs}
\end{equation}
Apart from computing the $T$'s by a variation of the matter Lagrangian wrt $M, C_i$ and $h_{ij}$, one can also obtain them via a decomposition of the usual energy momentum tensor $T^{\m\n}=\frac{1}{\sqrt{-g}}\frac{\delta \call_{\mathrm{mat}}}{\delta g_{\m\n}}$. Via \eqref{dgdecomp} one finds
\begin{equation}
T^0=T^{\m\n}n_\m n_\n \, ,\qquad T^i=T^{\m\n}n_{(\m}e_{\n)}^i \, ,\qquad T^{ij}= T^{\m\n}e_\m^i e_\n^j\,.\label{EMDecomp}
\end{equation}
By comparing \eqref{EMDecomp} to \eqref{EinstDecomp} the equations \eqref{Eeqs} then follow immediately. One advantage of the form \eqref{EMDecomp} is that it can also be used in cases where the matter has no Lagrangian description and is only defined in terms of an energy momentum tensor $T^{\m\n}$. Clearly the conservation of energy momentum $\nabla_\mu T^{\m\n}=0$ will become equivalent to some equations for the $T$'s. These equations, and their origin in diffeomorphism invariance, will be discussed in the next subsection.

\subsection{Diffeomorphism symmetries and Noether identities}
Because a choice of KS coordinates amounts to a choice of preferred time coordinate, it breaks manifest diffeomorphism invariance. Since they leave the time coordinate invariant, spatial diffeomorphisms, i.e. $\tilde x^i(x^j)$, do remain manifest. But more general coordinate transformations map one choice of KS coordinates into another; this will leave the theory invariant, but in a less manifest way. One reason to have a closer look at diffeomorphism invariance in the KS formulation is that via the associated Noether identities it allows to find the equivalent of the Bianchi identity and energy-momentum conservation of the standard covariant formulation. Of course these equations can also be obtained in more direct fashion via the decompositions (\ref{EinstDecomp}, \ref{EMDecomp}). A second reason is that invariance under these symmetries will explain much of the structure of the Lagrangian and the modified derivatives appearing. 

It will be useful to split the $d+1$ dimensional diffeomorphisms into two classes, which we will refer to as {\it time redefinitions} and {\it time-dependent spatial diffeomorphisms}. This split originates in the split of an infinitesimal diffeomorphism $\xi^\m$ into $\Lambda=\xi^\m n_\m$ and $\xi^i=\x^\mu e_\m^i$, via the KS frames (\ref{basexp2}, \ref{basexp2a}).

Although it is customary to discuss diffeomorphisms in the passive formulation, where one keeps the coordinates fixed and the action on tensors is via the Lie derivative, we find in this case the active formulation to be more insightful. For convenience we remind the reader of the relation between infinitesimal passive and active transformations and their action on an arbitrary tensor $S$:
\begin{align}
\!\!\!\!\!\delta_\mathrm{A}x^\mu&=\tilde x^\mu-x^\mu= -\xi^\m&\delta_\mathrm{A} S^\m_\n&=\tilde S^\m_\n(\tilde x)-S^\m_\n(x)=-\partial_\l\xi^\mu S^\l_\n+\partial_\n \xi^\l S^\m_\l \, ,\label{activetopassive1}\\
\!\!\!\delta_\mathrm{P}x^\mu&=0&\delta_\mathrm{P} S^\m_\n&=\tilde S^\m_\n(x)-S^\m_\n(x)=\delta_A S^\m_\n-\partial_\l S^\m_\n\delta_\mathrm{A}x^\l=L_\xi S^\m_\n \label{activetopassive2} .
\end{align}
It might be relevant to note how derivatives transform differently, i.e. $\delta_\mathrm{P}\partial_\m=0$, so that $\delta_\mathrm{P}\partial_\l S^\m_\n=\partial_\l \delta_{\mathrm{P}} S^\m_\n$, but $\delta_\mathrm{A}\partial_\m=\partial_\m \xi^\n\partial_\n$ so that 
\begin{equation}
\delta_{\mathrm{A}}\partial_\l S^\m_\n=\partial_\l\delta_{\mathrm{A}} S^\m_\n+\partial_\l \xi^\r \partial_\r S^\m_\n \,.
\end{equation}
The above generalizes straightforwardly to tensors with an arbitrary number of upper and lower indices. In the remainder of this section all infinitesimal transformations will be active and we drop the A subscript.

\subsubsection{Time redefinitions}
First we consider redefinitions of the coordinate $t$, i.e. diffeomorphisms of the form $\xi^\m n_\m=-M\Lambda$, $\xi^\mu e_\m^i=0$. The coordinates transform as
\begin{align}
\delta_\Lambda t&=-\Lambda(x,t)\, ,\\
\delta_\Lambda x^i&=0 \, .
\end{align}
Demanding the line element $ds^2$ -- see \eqref{metcase2} -- to be form invariant, leads to the transformations 
\begin{align}
\delta_\Lambda (-M^{-1})&=\hat{\partial}_t\Lambda \, ,\label{timred1}\\
\delta_\Lambda C_i&=\hat\partial_i\Lambda \, ,\\
\delta_\Lambda h_{ij}&=0\, .\label{timred3}
\end{align}
Note the strong similarity to the U(1) gauge transformations of Maxwell theory, as this will provide some intuition in the discussion of invariants below. We should stress however that this is but a similarity rather than a real equality, since contrary to standard U(1) gauge transformations here also the coordinates transform, the transformations are not abelian and it is the hatted rather than standard partial derivatives that appear in (\ref{timred1}-\ref{timred3}).

A short computation reveals that the following objects are invariant under these time redefinitions, i.e. $\delta_\Lambda X=0$, for $X$ any of
\begin{equation}
h_{ij}\,,\ \hat \partial_t\,,\ \hat \partial_i\,,\ K_{[ij]}\,,\ K_{(ij)}\,,\ \Omega_i\,,\ \hat \Gamma_{ij}^k\,,\ \hat R_{ij}\,,\ Mdtd^dx \, .\label{invlist}
\end{equation}
First of all, one sees that although the partial derivatives are not invariant\footnote{Remember we are discussing active transformations.}, their hatted versions are. So we see that the appearance of these hatted partial derivatives is no coincidence, but fully dictated by the invariance under time redefinitions. Since $\hat \Gamma_{ij}^k$ is built out of $h_{ij}$ and $\hat \partial_i$, it also follows that it is invariant, which makes it the natural connection to use. Similarly it is $\hat R_{ij}$, the modified Ricci tensor which is invariant rather than the standard Ricci tensor $R_{ij}$. The symmetric part $K_{(ij)}=\frac{1}{2}\hat{\partial}_t h_{ij}$ is manifestly invariant while the anti-symmetric part takes a form similar to a U(1) curvature tensor $K_{[ij]}=-M\hat{\partial}_{[i}C_{j]}$.

Given the list of invariants \eqref{invlist}, the KS Lagrangian \eqref{KSlag} is manifestly invariant. The same is true for the equations of motion by inspection of (\ref{G0}-\ref{Gij}). More generally speaking, time redefinition invariance guarantees that $M$, $C_i$ and derivatives can only appear through the objects \eqref{invlist}. 

\subsubsection{Time dependent spatial diffeomorphisms}
A second set of diffeomorphisms are those for which the generating vector field satisfies $\xi^\m n_\m=0$. The remaining components, $\xi^i=e^i_\m\xi^\m$, define a spatial coordinate transformation, but one that is time dependent. Explicitly
\begin{align}
\delta_\xi t&=0 \, ,\\
\delta_\xi x^i&=-\xi^i(x,t) \, .
\end{align}

It will be useful to introduce a separate notation for the time derivative of this vector field:
\begin{equation}
\hat f_i  =h_{ij}\hat\partial_t \xi^j\, . \\
\end{equation}
Invariance of $ds^2$ then defines the transformations of the fields:
\begin{align}
\delta_\xi (-M)^{-1}&=C_i\hat f^i\, ,\label{tdsd1}\\
\delta_\xi C_i&=\calc_\xi C_i-M^{-1}\hat f_i\, ,\\
\delta_\xi h_{ij}&=\calc_\xi h_{ij} \, .\label{tdsd3}
\end{align}
Here we introduced the following linear operation on tensor components:
\begin{align}
\calc_\xi S^{i_1\ldots i_n}_{j_1\ldots j_m}=&-S^{ki_2\ldots i_n}_{j_1\ldots j_m}\hat{\partial}_k\xi^{i_1}-\ldots-S^{i_1\ldots i_{n-1} k}_{j_1\ldots j_m}\hat{\partial}_k\xi^{i_n}\nonumber\\
&+S^{i_1\ldots i_n}_{kj_2\ldots j_m}\hat{\partial}_{j_1}\xi^k+\ldots+S^{i_1\ldots i_n}_{j_1\ldots j_{m-1}k}\hat{\partial}_{j_m}\xi^k \, . 
\end{align}
The introduction of $\calc_\xi$ is useful when performing variations of objects build out of the basic fields, since this operation is nicely compatible with tensor multiplication and contraction. E.g.
\begin{equation}
\calc_\xi(T^{i_1\ldots i_n}_{j_1\ldots j_m}S^{j_1\ldots j_m}_{k_1\ldots k_p})=\calc_\xi (T^{i_1\ldots i_n}_{j_1\ldots j_m})\calc_\xi(S^{j_1\ldots j_m}_{k_1\ldots k_p}) \, . 
\end{equation}
The KS action \eqref{KSlag} is not manifestly invariant under the transformations (\ref{tdsd1}-\ref{tdsd3}), a somewhat tedious calculation reveals that
\begin{equation}
\delta_\xi\left[Mdt\sqrt{h}d^d x\left(\hat R+K_{ij}K^{ij}-K^2\right)\right]=4Mdt\sqrt{h}d^d x\big(\hat D_i+\Omega_i\big)\left(\hat f_j K^{ji}-\hat f^iK\right) .
\end{equation}
It is then important to observe that
\begin{equation}
\label{coolformula}
(\hat D_i+\Omega_i)S^{[i j_1\ldots j_n]}=\frac{1}{M\sqrt{h}}\partial_i \left( M\sqrt{h}S^{[i j_1\ldots j_n]}\right)-\frac{1}{M\sqrt{h}}\partial_t \left(M\sqrt{h}\,C_i S^{[i j_1\ldots j_n]}\right)
\end{equation}
which implies
\begin{equation}
\delta_\xi S_{\mathrm{KS}}=\int dt d^d x \left[\partial_i\left(M\sqrt{h}(\hat f_j K^{ji}-\hat f^iK)\right)-\partial_t\left(M\sqrt{h}C_i(\hat f_j K^{ji}-\hat f^iK)\right) \right]=0
\end{equation}
This explicitly confirms that indeed general relativity in KS form is invariant also under time-dependent spatial diffeomorphisms, albeit not manifestly so. We stress that this is nothing but a consistency check, since the KS action is identical to the Einstein-Hilbert action and the time-dependent spatial diffeomorphisms are nothing but a subclass of $d+1$ dimensional diffeomorphisms, so invariance is guaranteed by construction.

\subsubsection{Noether identities}\label{noethid}
In any theory with a gauge or local symmetry the equations of motion are not all independent and the (differential) relations between them go under the name of Noether identities (sometimes also called classical Ward identities), see e.g. \cite{Avery:2015rga} for a pedagogic introduction. In the case of general relativity the Noether identities amount to the Bianchi identity for the Einstein tensor and the conservation equation for the energy-momentum tensor. 

Given a local symmetry of the form\footnote{Also here we take the active point of view}
\begin{equation}
\delta x^\mu=H^\m \lambda\, , \qquad \delta\phi=F(\phi,\partial\phi)\lambda+F^\mu(\phi)\partial_\m \l \, ,
\end{equation}
the associated Noether identity is
\begin{equation}
(F-H^\m \partial_\m\phi)\cale=\partial_\m(F^\m \cale) \,. 
\end{equation}
Here $\cale$ are the Euler-Lagrange equations, defined through a generic variation of the Lagrangian as
\begin{equation}
\delta \call=-\cale \delta\phi+\partial_\m\theta^\mu  \, .
\end{equation}

We can then apply this to the Lagrangian $\call_{\mathrm{KS}}(M,C,h)$ \eqref{KSlag} both for the time redefinitions and time dependent spatial diffeomorphisms discussed above. The resulting Noether identities are respectively equivalent to
\begin{align}
\hat D_i G^i+2\Omega_i G^i-\hat\partial_t G^0-KG^0-K_{ij}G^{ij}&=0 \, ,\\
\hat D_j G^{ji}+\Omega_jG^{ji}-\hat{\partial}_t G^i-2 G^jK_j{}^i-KG^i+\Omega^iG^0&=0 \, .
\end{align}
Of course the same equations can also be obtained by decomposing the Bianchi identity $\nabla_\m G^{\m\n}=0$ via the methods of the previous section. Through the Einstein equations in the form \eqref{Eeqs} one then directly obtains the energy-momentum conservation equations:
\begin{align}
\hat D_i T^i+2\Omega_i T^i-\hat\partial_t T^0-KT^0-K_{ij}T^{ij}&=0 \, ,\\
\hat D_j T^{ji}+\Omega_jT^{ji}+\Omega^iT^0-\hat{\partial}_t T^i-2 T^jK_j{}^i-KT^i&=0 \, .
\end{align}

\subsection{Conformal redefinition and the relativistic Poisson equation} \label{Poissec}
The field content of GR in KS formulation, as we introduced it above, is the triple $(M,C_i, h_{ij})$. But since $M$ appears algebraically in the Lagrangian its equation of motion, eqn \eqref{G0}, is a constraint and does not involve any derivatives on $M$. Second derivatives on $M$ do appear in eqn \eqref{Gij} but this is a tensorial equation instead of a scalar equation. This suggests that $M$ is not a relativistic analog of the Newtonian potential, nor is \eqref{G0} the analog of the Poisson equation. Indeed in a non-relativistic expansion, see section \ref{evsec} or \cite{Kol:2010si, Hansen:2020pqs}, it is the combination $G^0+G^{ij}h_{ij}$ that becomes the Newtonian Poisson equation.  This particular linear combination appears naturally as the equation of motion $E^\psi$ of a scalar $\psi$ introduced through the field redefinition\footnote{Let us point out that from here onward in the paper we set the number of spatial dimensions $d=3$.}
\begin{equation}
M=e^\frac{\psi}{2}\, ,\qquad h_{ij}=e^{-\psi}\g_{ij}\, .\label{fieldredef}
\end{equation}
Indeed, defining\footnote{Throughout the remainder of the paper we will use the notation $\call = \sqrt \g L$ and $\cale = \sqrt \g E$ to distinguish between tensorial quantities and the associated densities.}
\begin{equation}
\delta \call_{\mathrm{KS}}=\sqrt{\g}\left(E^\psi \delta\psi-2E^i\delta C_i-E^{ij}\delta \g_{ij}\right)+\partial_\m\tilde\theta^\m \, ,\label{varKSW}
\end{equation}
a comparison to \eqref{vargrav} reveals that
\begin{equation}
E^\psi=e^{-\psi}(G^0+h_{ij}G^{ij}),\qquad E^i=e^{-\psi/2}G^i,\qquad E^{ij}=e^{-2\psi}G^{ij}\,.
\end{equation}

This suggests $\psi$ is a natural relativistic analog of the Newtonian potential, and indeed this change of variables has proven very fruitful in the nonrelativistic expansion; this was the key insight of \cite{Kol:2007bc} that motivated the KS formulation of \cite{Kol:2010si} and it also appeared in the $1/c$ expansion in \cite{Hansen:2020pqs, Ergen:2020yop}.

Expressed in terms of the alternative fields \eqref{fieldredef}, and upon dropping some total derivative terms, the KS action \eqref{KSlag} becomes
\begin{equation}
S_{\mathrm{KS}} = \int dt d^3 x \sqrt \g \Big( \hat R + e^{-2\psi} (\tilde K_{(ij)} \tilde K^{ij}  - \tilde K^2) + e^{2\psi} \tilde K_{[ij]} \tilde K^{ij} - \frac{1}{2}\hat \partial_i \psi\hat \partial^i \psi -  2 \tilde \O^i\hat \partial_i \psi\Big)\, ,\label{KSW}
\end{equation}
where now
\begin{equation}
\hat R=\hat R[\g]=\g^{ij}\hat{R}_{ij}[\g]\quad\mbox{and}\quad \hat{R}_{ij}[\g]=\hat\partial_{k}\hat\G_{ij}^k[\g]-\hat\partial_{j}\hat\G_{ik}^k[\g]+\hat\G_{lk}^l[\g]\hat\G_{ij}^k[\g]-\hat\G_{ik}^l[\g]\hat\G_{lj}^k[\g] \, ,
\end{equation}
with $\Gamma_{ij}^k[\g]$ the Levi-Civita connection of $\g_{ij}$. Furthermore
\begin{equation}
\tilde K_{ij} = e^{\psi}K_{ij}=\frac{1}{2}(\partial_t \g_{ij} - \g_{ij} \partial_t \psi ) - \hat\partial_{[i}C_{j]}\, , \qquad \tilde \Omega_i=-\partial_t C_i\,.
\end{equation}
Note that if one interprets $\psi$ as the relativistic analog of the Newtonian potential then $\Omega^i = \frac{1}{2}\hat \partial^i \psi + \tilde \O^i $ is a natural analog of the gravitational force vector.

\subsection{Making time derivatives explicit}\label{expltimesec}
General relativity in the KS formulation appears -- at least to us -- most elegant in the form \eqref{KSlag}, which is simple and very similar to the ADM form. In the previous subsection we made a change of variables that is more suitable for comparison to Newtonian gravity. This introduces a few extra terms in the Lagrangian, see \eqref{KSW}, but still it retains some of its elegance due to the use of the quantities $\hat R$, $\tilde K$ and the hatted partial derivative $\hat \partial_i$ (see \eqref{hpKS}). These quantities each contain 'hidden' time derivatives, which we would now like to make explicit. This is motivated by our aim, in sections \ref{expsec} and \ref{evsec}, to make an expansion in (inverse) powers of $c$, which naturally accompany each time derivative. The key point in this subsection is that in particular the time dependence in the term $\sqrt{\g}\hat R$ can be simplified quite a lot by discarding a total derivative term. 

We start by introducing the speed of light $c$ in the original relativistic metric \eqref{metcase2}, which together with the redefinition \eqref{fieldredef} is then of the form
\begin{equation}
ds^2=-e^{\psi}(c\,dt+C_i dx^i)^2+e^{-\psi}\g_{ij}dx^i dx^j\,.\label{metex}
\end{equation}
Since apart from an overall factor in front of the action this is the only place $c$ appears in the purely metric sector of the theory it implies all factors of $c$ can always be re-absorbed by simply rescaling $t$. Equivalently this implies that each time derivative has to be accompanied by a power $c^{-1}$. Since the Lagrangian is maximally second order in derivatives it implies we can write
\begin{equation}
\call=\sqrt{\g}\left(L_0+c^{-1}L_1+c^{-2}L_2\right)\, ,\label{totlag}
\end{equation}
with $L_0$ containing no time derivatives, $L_1$ being first order in time derivatives and $L_2$ second order. We will discard total derivatives such that $L_1$ and $L_2$ will be quadratic in derivatives rather than contain second derivatives. $L_0$ will still contain second spatial derivatives. One can think of $L_2$ as the kinetic term, of $L_1$ as a Lorentz coupling to velocity and of $L_0$ as the potential term.

To simplify notation we will introduce a dot notation for time derivatives, i.e. $\dot{{}}=\partial_t$\ . We make the important conventional choice that {\it the time derivative will always act on tensors with lower indices}. This implies that if one sees an object with upper indices and a dot, one should first lower the indices before interpreting the dot as a time derivative! For example:
\begin{equation}
\dot C_i=\partial_t C_i\ \mbox{ and }\ \dot C^i=\g^{ij}\dot C_j=\g^{ij}\partial_t C_j\,,\ \mbox{while }\ \partial_t C^i=(\partial_t\gamma^{ij}) C_j+\dot C^i\neq \dot C^i \, ,
\end{equation}
or
\begin{equation}
\dot \g_{ij}=\partial_t \g_{ij}\ \mbox{ and }\ \dot \g^{ij}=\g^{ik}\g^{kl}\dot \g_{ij}=\g^{ik}\g^{kl}\partial_t \g_{ij}\,,\mbox{ while }\ \partial_t \g^{ij}=-\dot \g^{ij}\neq \dot \g^{ij} \, .
\end{equation}
Note that under our convention we also have
\begin{equation}
\dot \g=\g^{ij}\dot \g_{ij}=\g^{ij}\partial_t \g_{ij}=\partial_t\log\det (\g_{ij}) \, ,
\end{equation}
One could proceed by simply inserting \eqref{metex} into the Einstein Hilbert action, but we can use the results of the previous subsections to start from \eqref{KSW} instead. First we introduce the appropriate factors of $c$ and split all the objects appearing in \eqref{KSW} in parts  containing equal powers of $c$:
\begin{align}
\tilde K_{(ij)}&=\frac{c^{-1}}{2}(\dot \g_{ij}-\g_{ij}\dot \psi) \, ,\label{spl1}\\
\tilde K_{[ij]}&=-\frac{1}{2}C_{ij}+c^{-1}C_{[i}\dot C_{j]},\qquad\quad C_{ij}=\partial_i C_j-\partial_j C_i \, ,\\
\tilde \Omega_i&= -c^{-1}\dot C_i \, ,\\
\hat{\partial}_i\psi&=\partial_i \psi-c^{-1}C_i\dot \psi \,. \label{spl4}
\end{align}
In rewriting $\hat R[\gamma]$ it is useful to isolate a total derivative (the second line below), which will drop out of the action. One calculates that
\begin{align}
\hat R=&\,R+ c^{-1}\Gu^{ij\,kl}\dot \g_{ij}D_kC_l+c^{-2}\left(\Gu^{ij\,kl}\dot C_i C_j\dot \g_{kl}-\frac{1}{4} \Gu^{ij\,kl\,mn}C_iC_j\dot \g_{kl}\dot \g_{mn}\right)\nonumber\\
&-2c^{-1}D_i\left(\Gu^{ij\,kl}C_j\dot \g_{kl}\right)+c^{-2}D_t\left(\Gu^{ij\,kl}C_iC_j\dot \g_{kl}\right)\, , \label{Rh}
\end{align}
where $D_i$ and $R$ are defined with respect to the Levi-Civita connection of $\g_{ij}$ and we additionally introduced
\begin{align}
D_t\cdot &=\frac{1}{\sqrt{\g}}\partial_t(\sqrt{\g}\,\cdot)=\partial_t\cdot+\frac{1}{2}\dot{\gamma}\cdot \, , \\
\Gu^{ij\,kl}&=-\g^{ij}\g^{kl}+\g^{il}\g^{kj} \, , \\
\Gu^{ij\,kl\,mn}&=\g^{ij}(\g^{kl}\g^{mn}-\g^{kn}\g^{ml})+2\g^{in}(\g^{kj}\g^{ml}-\g^{kl}\g^{mj}) \, .
\end{align}
Inserting (\ref{spl1}-\ref{spl4}) and \eqref{Rh} into \eqref{KSW} leads, upon dropping total derivatives, to
\begin{align}
	L_0=&\,R+\frac{1}{4}e^{2\psi}C_{ij}C^{ij}-\frac{1}{2}\partial_i\psi\partial^i\psi \, ,\nonumber\\
	L_1=&\, \Gu^{ij\,kl}D_iC_j\dot \g_{kl}+(2\partial^j\psi-e^{2\psi}C^{ij}C_i)\dot C_j+C^i\partial_i \psi\dot \psi \, ,\nonumber\\
	L_2=&\,-\Gu^{ij\,kl}C_i\dot C_j\dot \g_{kl}+e^{-2\psi}\dot \g \dot \psi -2C^i \dot C_i\dot\psi\label{L012}\\
	&+\frac{1}{4}(e^{-2\psi}\Gu^{kl\,mn} +\Gu^{ij\,kl\,mn}C_iC_j)\dot \g_{kl}\dot \g_{mn}-\frac{1}{2}e^{2\psi}\Gu^{ij\,kl}C_iC_j\dot C_k\dot C_l\nonumber\\
	&-\frac{1}{2}( 3e^{-2\psi}+C_i C^i) \dot{\psi}^2 \, .\nonumber
\end{align}
Let us stress again that the Lagrangian (\ref{totlag}, \ref{L012}) equals the Einstein-Hilbert Lagrangian up to a total derivative. We refer to Appendix \ref{eomapp} for the equations of motion in a form with explicit time derivatives. These can be obtained by varying (\ref{totlag}, \ref{L012}) or equivalently by rewriting (\ref{G0}-\ref{Gij}).

We will refrain from an extensive discussion of the symmetries in this form, as we already discussed them in some detail in a more compact form in section \ref{secsym}. The discussion there also implies the transformations \eqref{tred} and \eqref{texdsd} do indeed leave the action for (\ref{totlag}, \ref{L012}) invariant. Still let us shortly mention a few key points and formulas. For use in the next sections we find it convenient to revert in this and the next sections to the passive formulation.
In the passive formulation, using (\ref{activetopassive1},\ref{activetopassive2}),  and upon \eqref{fieldredef},  the time redefinition symmetry (\ref{timred1}-\ref{timred3}) takes the form 
\begin{align}
\delta_\Lambda \psi&=c^{-1}(\Lambda\dot \psi+2\dot \Lambda) \, ,\nonumber\\
\delta_\Lambda C_i &=\partial_i\Lambda+c^{-1}(\Lambda\dot{C}_i-\dot{\Lambda}C_i)\, ,\label{tred}\\
\delta_\Lambda \g_{ij}&=c^{-1}(\Lambda\dot{\g}_{ij}+2\dot{\Lambda}\g_{ij}) \, .\nonumber
\end{align}
while the time-dependent spatial diffeomorphisms (\ref{tdsd1}-\ref{tdsd3}) become
\begin{align}
\delta_\xi \psi&=L_\xi\psi+2c^{-1}C_i\partial_t\xi^i \, ,\nonumber\\
\delta_\xi C_i&=L_\xi C_i-c^{-1}(e^{-2\psi}\gamma_{ij}+C_iC_j)\partial_t \xi^j\, ,\label{texdsd}\\
\delta_\xi\gamma_{ij}&=L_\xi \gamma_{ij}+2c^{-1}(C_k\gamma_{ij} - C_{(i}\gamma_{j)k})\partial_t\xi^k \, .\nonumber
\end{align}

One point of importance below is that the time-redefinition symmetry \eqref{tred} implies a simpler scaling symmetry. Consider the special case $\Lambda=c(t-t_0)\alpha$, with $t_0$ an arbitrary constant and $\alpha$ a real parameter. Then observe that \eqref{fieldredef} implies that, when evaluated at $t=t_0$, the transformations of the fields and their time derivatives take the scaling forms $\delta_\alpha X=[X] X$, $\delta_\alpha \dot X=\alpha ([X]+1)\dot X$, where the scaling weights are
\begin{equation}
\!\!\!\!\!\!\!\! [\partial_i]=[\partial_i\psi]=0\,, \,\,[\partial_t]=[\partial_t\psi]=-[C_i]=1\,, \,\,[e^\psi]=-[e^{-\psi}]=[\gamma_{ij}]=-[\gamma^{ij}]=2\, .\label{tweights}
\end{equation}
From this it follows that $[\sqrt{\gamma}dtd^3x]=2$, which in turn implies that invariance of the action requires $[L]=-2$. One can indeed check that with the weights \eqref{tweights} all the terms in \eqref{L012} carry weight $-2$. Note also the simple fact that since only $\partial_t$ and $C_i$ have an odd weight, every term with a single time derivative should carry an odd number of $C_i$ factors, something which indeed can also be verified by inspection of \eqref{L012}. This feature will play a small but key role in the shuffling algorithm introduced in section \ref{shufflingalg} and it is thus interesting to point out it has its origin in the time redefinition symmetry of the theory.

The time redefinition symmetry \eqref{tred} will also play a key role in gauge fixing the trace of the subleading metric coefficients, as discussed in section \ref{tracesec}. In section \ref{evexplsec} we will shortly discuss how the symmetries (\ref{tred}, \ref{texdsd}) upon expansion lead to the non-relativistic symmetries of \cite{VandenBleeken:2017rij, Hansen:2019pkl, Hansen:2020pqs}.

\section{The $1/c$ expansion in KS formalism}\label{expsec}
For a pedagogic summary and overview of earlier work on the $1/c$ expansion see \cite{Hansen:2021xhm, Toappearreview}. All of this work has been in a form making $4$-dimensional coordinate invariance manifest, and for that reason the analysis was based on Newton-Cartan geometry. Here we will revisit this expansion, but now in a 3+1 formulation, i.e. one that only keeps $3$-dimensional coordinate invariance manifest and uses an explicit choice of time. This is not  unnatural for two related reasons. First of all in the non-relativistic expansion there is a time direction that all non-relativistic observers agree upon. Secondly, by dimensional reasons, all time derivatives are accompanied by a power of $1/c$ and so the $1/c$ expansion is actually an expansion in time derivatives. As we argued in section \ref{intro}, it is the KS formalism that provides the natural 3+1 split to use for the $1/c$ expansion. In this 3+1 form the general structure of the expansion will be more transparent, which allows us to make some new all order observations as well as push the expansion to higher order than was previously done. In section \ref{evsec} we discuss the truncation of the $1/c$ expansion to the $1/c^2$ expansion in detail.

\subsection{Structure of the expansion}
The starting point is the relativistic KS Lagrangian in the form (\ref{totlag}, \ref{L012}). To perform the expansion we assume\footnote{A priori one could consider a Laurent expansion in $1/c$, a transseries including terms of the form $e^{-\frac{A}{c}}$ or even more generic $c$ dependence in the fields. Such more general ansatze remain largely unexplored and fall outside the scope of this paper.} the fields to be analytic in $1/c$\,:
\begin{equation}
\psi=\sum_{n=0}^\infty \os{\psi}{n}c^{-n}\qquad C_i=\sum_{n=0}^\infty \os{C}{n}_ic^{-n}\qquad \gamma_{ij}=\sum_{n=0}^\infty \os{\gamma}{n}_{ij}c^{-n} \, .\label{expansans}
\end{equation}
Note that the Lagrangian (\ref{totlag}, \ref{L012}) has some explicit dependence on $1/c$ as well. In combination with the $1/c$ dependence of the fields \eqref{expansans} one gets an expansion
\begin{equation}
\call[\psi,C_i,\gamma_{ij};c]=\sum_{n=0}^\infty \os{\call}{n}[\os{\psi}{0},\os{C}{0}_i,\os{\gamma}{0}_{ij},\ldots, \os{\psi}{n},\os{C}{n}_i,\os{\gamma}{n}_{ij}]c^{-n} \, .\label{lagans}
\end{equation}
In the following subsections we will discuss some general features of this expansion. It will often be useful to collect all fields and their expansion coefficients in a single 'master field':
\begin{equation}
\Phi=(\psi,C_i,\gamma_{ij})\,,\qquad \Phi=\sum_{n=0}^\infty \os{\Phi}{n}c^{-n}\,.
\end{equation}

\subsubsection{Actions and equations of motion}\label{eomstrucsec}
Note that one could either expand the Lagrangian $\call$ and then vary the expansion coefficients $\os{\call}{n}$ with respect to the various expanded fields to obtain equations of motion, or alternatively one could expand the original equations of motion associated to the unexpanded Lagrangian. In \cite{Hansen:2020pqs} the interplay and compatibility between these two approaches was explained, for completeness we shortly revisit this in appendix \ref{acteomapp}. The key relation derived there is
\begin{equation}
\frac{\delta \os{S}{n+m}}{\delta \os{\Phi}{m}}=\os{\cale}{n}\, .\label{exprelsmain}
\end{equation}
In words: the $n$'th order equations of motion can be obtained by varying the $(n+m)$'th order action with respect to the $m$'th order fields.
The relations \eqref{exprelsmain} can be interpreted in two ways. One point of view is that to work up to $n$'th order we can simply use the action $\os{S}{n}[\os{\Phi}{0},\ldots,\os{\Phi}{n}]$, which can reproduce all equations of motion up to $n$'th order by variation with respect to the fields $\os{\Phi}{k}$, $k=0,\ldots, n$. I.e. one could expand the action up to the preferred order $n$ and then vary it with respect to all fields up to that order, to obtain all the relevant equations of motion: 
\begin{eqnarray}
\os{\cale}{k}=\frac{\delta \os{S}{n}}{\delta \os{\Phi}{n-k}}\qquad k\leq n \, .
\end{eqnarray} 
In practice this approach is however a rather contrived way to find, say, $\os{\cale}{1}$, which one could also obtain by varying $\os{S}{1}$ with respect to $\os{\Phi}{0}$, a much shorter calculation. Indeed from a computational point of view it is more natural to compute order by order:
\begin{equation}
\os{\cale}{k}=\frac{\delta \os{S}{k}}{\delta \os{\Phi}{0}}\qquad k\leq n \, .\label{compscheme}
\end{equation}
I.e. one can obtain all equations of motion up to a given order $n$, by expanding the action up to order $n$ and then varying all of these actions only with respect to the leading order fields $\os{\Phi}{0}$. As we will discuss in the next subsection, section \ref{tracesec}, this approach has the additional advantage that if we restrict only to variations with respect to the leading order fields $\os{\Phi}{0}$ then the action $\os{S}{k}$ can be replaced by an equivalent but simpler action $\os{\bar{S}}{k}$.
 
\subsubsection{Gauge fixing the trace}\label{tracesec}
The non-relativistic theory obtained through the $1/c$ expansion contains a tower $\os{\gamma}{n}_{ij}$ of symmetric tensor fields. It is natural to identify $\os{\gamma}{0}_{ij}$ as a metric defining the geometry underlying the expanded theory. Using this metric we can split the subleading tensor fields into a trace and traceless part\footnote{While $\os{\gamma}{n}$ indicates the {\it trace} of $\os{\g}{n}_{ij}$ for $n\geq 1$, we will use $\os{\gamma}{0}$ to indicate the {\it determinant} of $\os{\gamma}{0}_{ij}$. Although this might at first appear confusing this will not lead to any clash of notation as the trace of $\os{\gamma}{0}_{ij}$ will never appear (since $\os{\gamma}{0}^{ij}\os{\gamma}{0}_{ij}=3$) and neither will the determinants of  $\os{\g}{n}_{ij}$ (since the expansion of $\det \gamma_{ij}$ can be expressed in terms of $\det \os{\g}{0}_{ij}$ and traces of polynomials in $\os{\g}{k}_{ij}$). While the distinctive interpretation of $\os{\gamma}{0}$ versus $\os{\gamma}{n}$, $n\geq 1$, puts extra load on the notation, it prevents the proliferation of symbols, which is already quite large.}:
\begin{equation}
\os{\gamma}{n}=\os{\gamma}{0}^{ij}\os{\gamma}{n}_{ij}\, ,\qquad\os{\bar \gamma}{n}_{ij}=\os{\gamma}{n}_{ij}-\frac{1}{3}\os{\gamma}{n}\os{\gamma}{0}_{ij}\, ,\qquad n\geq  1 \, .\label{tracesplit}
\end{equation}
The trace appears frequently in the expansion, both in the equations of motion and the Lagrangian, for example through the expansion of the determinant of $\gamma_{ij}$. As we discuss in detail in appendix \ref{techtraceapp}, one can always go to a coordinate gauge where all traces are zero, i.e. $\os{\gamma}{n}=0, \forall n\geq 1$. In such a gauge the equations of motion will simplify accordingly. Concerning the action one has to be a little more careful, since gauge fixing at the level of the action can lead to a loss of equations of motion\footnote{These are the so called constraints, such as the well known Gauss constraint of electro-magnetism that needs to be supplemented to the Euler-Lagrange equations obtained from the Lagrangian in $A_0=0$ gauge.}. This can however be circumvented in the $1/c$ expansion, in the following sense. Let us define
\begin{equation}
\os{\bar S}{n}=\os{S}{n}\big|_{\substack{\os{\gamma}{m}=0\\  m\geq 1}}\, ,\qquad \os{\bar\cale}{n}=\frac{\delta \os{\bar S}{n}}{\delta \os{\Phi}{0}} \, ,\label{gfvar}
\end{equation}
then it is shown in appendix \ref{techtraceapp} that
\begin{equation}
\left.\left\{\os{\cale}{0}=0,\ldots, \os{\cale}{n}=0\right\}\right|_{\substack{\os{\gamma}{m}=0\\  m\geq 1}}\quad\Leftrightarrow\quad \left\{\os{\bar\cale}{0}=0,\ldots, \os{\bar\cale}{n}=0\right\}\, .\label{eqeoms}
\end{equation}
With the above we mean that the gauge fixed collection of equations of motion is equivalent to the collection of equations of motion obtained from the action which has been gauge fixed. 

We should point out that the precise interpretation of \eqref{gfvar} is somewhat subtle when the variation with respect to $\os{\gamma}{0}_{ij}$ is considered. One can treat $\os{\bar{\g}}{n}_{ij}$ as an independent field that does not change under variation of $\os{\gamma}{0}_{ij}$, which is consistent if one does not forget to replace all expressions of the form $\bar\gamma_{ij}F^{ij}$ by $\bar{\gamma}_{ij}\bar F^{ij}$ before varying. Alternatively one can consider $\os{\bar{\g}}{n}_{ij}=\os{\g}{n}_{ij}-\frac{1}{3}\os{\gamma}{0}_{ij} \os{\gamma}{0}^{kl}\os{\g}{n}_{kl}$, keep $\os{\g}{n}_{ij}$ fixed while varying $\os{\gamma}{0}_{ij}$ and only after varying impose $\os{\gamma}{n}=0$. Both approaches are equivalent, as we spell out in appendix \ref{techtraceapp}.

\subsubsection{The leading order as the stationary sector of GR}\label{LOsec}
We have chosen notation and conventions in such a way that both the Lagrangian and the fields start at order $c^{0}$, see \eqref{totlag} and (\ref{expansans}, \ref{lagans}). It follows that the leading order of the $1/c$ expansion is
\begin{equation}
\os{\call}{0}=\sqrt{\os{\gamma}{0}}\os{L}{0}_0\,,
\end{equation}
where via \eqref{L012}
\begin{equation}
\os{L}{0}_0=R[\os{\gamma}{0}]+\frac{1}{4}e^{2\os{\psi}{0}}\os{C}{0}_{ij}\os{C}{0}^{ij}-\frac{1}{2}\partial_i\os{\psi}{0}\partial^i\os{\psi}{0}\,.
\end{equation}
In the above $R[\os{\gamma}{0}]$ is the Ricci tensor of the metric $\os{\gamma}{0}_{ij}$ and $\os{C}{0}_{ij}=\partial_i \os{C}{0}_{j}-\partial_j \os{C}{0}_{i} $, note that furthermore all indices are raised with $\os{\gamma}{0}^{ij}$, which is the inverse of $\os{\gamma}{0}_{ij}$.

Since the leading order fields $\os{\psi}{0}, \os{C}{0}_{i}, \os{\gamma}{0}_{ij}$ will appear at all lower orders as well, it will be useful to introduce notation removing their superscript. {\it By abuse of notation we will refer to the leading order fields in the expansion by exactly the same symbol as the un-expanded fields. This should not lead to confusion as it should be clear from the context whether we are discussing the $1/c$ expansion or not.} I.e. from here onwards:
\begin{equation}
\os{\psi}{0}\rightarrow \psi\,,\qquad \os{C}{0}_i\rightarrow C_i\,,\qquad \os{\gamma}{0}_{ij}\rightarrow\gamma_{ij}\,.
\end{equation}
Similarly we will simply write $\sqrt{\gamma}$, $R$ and $C_{ij}$ for $\sqrt{\os{\gamma}{0}}$, $R[\os{\gamma}{0}]$ and $\os{C}{0}_{ij}$ respectively. Additionally indices will be raised {\it at all orders of the expansion} with $\gamma^{ij}$, again abusive notation for $\os{\gamma}{0}^{ij}$. Similarly we will simply use $\Gamma_{ij}^k$ and $D_i$ for the Levi-Civita connection $\Gamma_{ij}^k[\os{\gamma}{0}]$ and covariant derivative $D_i[\os{\gamma}{0}]$.

In this new notation then, we can summarize the Lagrangian at leading order as

\begin{equation}
\os{\call}{0}=\sqrt{\gamma}\left(R+\frac{1}{4}e^{2\psi}C_{ij}C^{ij}-\frac{1}{2}\partial_i\psi\partial^i\psi\right)\,.\label{LOlag}
\end{equation}
This form immediately reveals that this leading order Lagrangian \eqref{LOlag} is nothing but the full KS Lagrangian (\ref{totlag}, \ref{L012}) with all time derivatives removed. In other words, for time independent field configurations the leading order $1/c$ expansion is exact. If one furthermore recalls the form of the metric \eqref{metex} then one sees that such time independent fields correspond exactly to a 4 dimensional stationary (lorentzian) metric. We can thus conclude that the leading order of the $1/c$ expansion captures the full non-linear dynamics of the stationary sector of general relativity. The same conclusion was reached in \cite{Ergen:2020yop}, but the approach taken in this paper has the advantage that it is much more straightforward to arrive at the leading order Lagrangian \eqref{LOlag} starting from the KS Lagrangian \eqref{totlag}, instead of using the fully covariant approach of \cite{Ergen:2020yop}.

Note that stationary solutions to GR, in addition to solving the dynamics described by \eqref{LOlag}, will also not source any higher order corrections. One can however consider quasi-stationary solutions to GR, i.e. stationary solutions in which one makes the integration constants time dependent. Then such metrics will still solve the dynamics of \eqref{LOlag}, since this does not contain any time derivatives, but the time dependent integration constants will lead to non-zero time derivatives which in turn will source subleading corrections. We discuss the structure of these subleading corrections in the following subsection.

\subsubsection{The universal linear part}\label{linsec}
As we discussed in the previous subsection, the leading order equations are non-linear equations for the leading order fields $\os{\Phi}{0}$. But once one considers the subleading equations, they are, as we will discuss in this subsection, linear equations determining the subleading fields $\os{\Phi}{n}, n\geq 1$. Furthermore these equations take the schematic form
\begin{equation}
\mathbb{D}^2 \os{\Phi}{n}=\os{\mathbb{S}}{n}[\os{\Phi}{n-1}, \ldots, \os{\Phi}{0}, \partial_t\!\!\os{\Phi}{n-1}, \ldots, \partial_t\os{\Phi}{0}, \partial_t^2\!\os{\Phi}{n-2}, \ldots, \partial_t^2\os{\Phi}{0} ] \, .\label{formlineq}
\end{equation}
Here $\mathbb{D}^2$ is a second order linear differential operator that is universal, i.e. it is the same at each order $n\geq 1$. The right hand side $\os{\mathbb{S}}{n}$ is a source term, built out of the fields of order lower than $n$, and their time derivatives. This gives the equations of motion in the $1/c$ expansion a hierarchic structure: at each order one determines $\os{\Phi}{n}$ by solving \eqref{formlineq}, and this field and its time derivatives then enter the source term in the equation for the higher order fields. 

To arrive at the equations in the form \eqref{formlineq} we first recall that the expansion of the Lagrangian takes the form:
\begin{equation}
\call=\sum_{n=0}^\infty \os{\call}{n}c^{-n} \, ,
\end{equation}
where, via \eqref{totlag},
\begin{equation}
\os{\call}{n}=\os{\call}{n}_0+\os{\call}{n-1}\!\!\!{}_1+\os{\call}{n-2}\!\!\!{}_2 \, .
\end{equation}
Now remark that the $n$'th order fields $\os{\Phi}{n}$ do not appear into the last two terms, but only in the first. Furthermore, since the fields $\os{\Phi}{n}$ by definition enter with a power $c^{-n}$  they will appear linearly in this first term, as $\os{\call}{n}$ comes with that same power. This first of all suggests to decompose
\begin{equation}
\os{\call}{n}=\os{\call}{n}_{\mathrm{lin}}+\os{\call}{n}_{\mathrm{source}} \, ,
\end{equation}
where
\begin{equation}
\os{\call}{n}_{\mathrm{source}}=\left.\os{\call}{n}_0\right|_{\os{\gamma}{n}_{ij}=0,\,\os{C}{n}_{i}=0,\, \os{\psi}{n}=0 }+\os{\call}{n-1}\!\!{}_1+\os{\call}{n-2}\!\!{}_2 \, .
\end{equation}
The second observation is then that the linear part is universal. One computes from \eqref{L012} that\footnote{We remind the reader that we have now started using notation where $\psi=\os{\psi}{0}$, etc. see section \ref{LOsec}. Furthermore we have gauge fixed the traces $\os{\gamma}{n}$ to zero, see section \ref{tracesec}, replacing $\os{\call}{n}$ by $\os{\bar{\call}}{n}$. For those readers interested in the results outside this gauge we have provided the additional terms proportional to the trace in appendix \ref{traceapp}.}
\begin{equation}
\os{\bar\call}{n}_{\mathrm{lin}}=\sqrt{\g}\Bigg(-\partial_i\psi\partial^i\os{\psi}{n}+\frac{\os{\psi}{n}e^{2\psi}}{2}C_{ij}C^{ij}+\frac{e^{2\psi}}{2}C^{ij} \os{C}{n}_{ij}-\os{\bar\g}{n}^{ij}(R_{ij}+\frac{e^{2\psi}}{2}C_{ik}C_{j}{}^k-\frac{1}{2}\partial_i\psi\partial_j\psi)\Bigg)\, .\label{linLag}
\end{equation}
The equations of motion then take the form \eqref{formlineq} with
\begin{equation}
\overline{\mathbb{D}}^2\os{\Phi}{n}=\frac{1}{\sqrt{\gamma}}\frac{\delta \os{\bar\call}{n}_\mathrm{lin}}{\delta \os{\Phi}{0}}\qquad\qquad \os{\overline{\mathbb{S}}}{n}=-\frac{1}{\sqrt{\g}}\frac{\delta \os{\bar\call}{n}_\mathrm{source}}{\delta \os{\Phi}{0}}\, .\label{lindefs}
\end{equation}

Since $\os{\bar\call}{n}_\mathrm{lin}$ has the same form $\eqref{linLag}$ at each order, $ \overline{\mathbb{D}}^2\os{\Phi}{n}$ needs to be computed only once. Doing so leads to
\begin{align}
\overline{\mathbb{D}}^2\os{\psi}{n}=&\,D_i\partial^i\os{\psi}{n}+\os{\psi}{n}e^{2\psi}C_{ij}C^{ij}+e^{2\psi}C^{ij}\os{C}{n}_{ij}+e^{2\psi}\os{\bar\g}{n}^{ij}C_{ik}C_{j}{}^k+\partial_i\psi D_j\os{\bar\g}{n}^{ij} \, ,\nonumber\\
\overline{\mathbb{D}}^2\os{C}{n}^j=& \,-D_i\left(e^{2\psi}(\os{C}{n}^{ij}+2\os{\psi}{n}C^{ij}+\os{\bar\g}{n}^{ik}C_k{}^j-\os{\bar\g}{n}^{jk}C_k{}^i)\right)\, ,\nonumber\\
\overline{\mathbb{D}}^2\os{\bar\g}{n}^{ij}=&\,\frac{1}{2}D^2 \os{\bar\g}{n}^{ij}+\partial^{(i}\psi\partial^{j)}\os{\psi}{n}-\os{\psi}{n}e^{2\psi}C^i{}_{k}C^{jk}-e^{2\psi}C^{(i}{}_k \os{C}{n}^{j)k}\label{DD2}\\
&+ \frac{1}{2}\os{\bar\g}{n}^{kl}e^{2\psi}C_{k}{}^iC_l{}^j-\frac{1}{2}\g^{ij}\bigg(\partial_k\psi\partial^k\os{\psi}{n}-\frac{\os{\psi}{n}}{2}e^{2\psi}C_{kl}C^{kl}-\frac{e^{2\psi}}{2}C^{kl} \os{C}{n}_{kl} \bigg)\nonumber\\
&+(\os{\bar\g}{n}^{ik}\g^{jl}+\os{\bar\g}{n}^{jk}\g^{il}-\frac{1}{2}\g^{ij}\os{\bar\g}{n}^{kl})(R_{kl}+\frac{e^{2\psi}}{2}C_{km}C_l{}^{m}-\frac{1}{2}\partial_k\psi\partial_l\psi)\nonumber\\
&- \frac{1}{3}\os{\bar\g}{n}^{ij} (R + \frac{e^{2\psi}}{2 }C_{kl}C^{kl} - \frac{1}{2}\partial_k \psi \partial^k \psi) \nonumber\, .
\end{align}
where
\begin{eqnarray}
	D^2 \os{\g}{n}^{ij} = D_k D^k \os{\g}{n}^{ij}-D_k D^i \os{\g}{n}^{jk}-D_k D^j\os{\g}{n}^{ik}+\g^{ij}D_k D_l \os{\g}{n}^{kl}\, .\label{D2}
\end{eqnarray}

Contrary to $\overline{\mathbb{D}}^2$, the source term $\os{\overline{\mathbb{S}}}{n}$ is different at each order -- indeed it becomes more intricate at each successive order -- and thus it needs to be computed for each order separately via \eqref{lindefs}. We provide the results for $\os{\bar\call}{1}_\mathrm{source}$ and $\os{\bar\call}{2}_\mathrm{source}$ in section \ref{explicitsec}.

\subsection{Explicit expansion up to NNLO}\label{explicitsec}
We discussed the leading order Lagrangian in section \ref{LOsec} and the all order linear part of the Lagrangian and equations of motion in section \ref{linsec}. Now we will present the complete Lagrangians at the first two subleading orders, in the traceless gauge -- see section \ref{tracesec}. The additional terms that appear without this gauge choice can be found in appendix \ref{traceapp}.

For notational convenience we have decided to remove the superscripts indicating the order and instead use different characters to label the different coefficients\footnote{As before, see section \ref{LOsec}, we indicate the leading order field with the same symbol as the complete $c$ dependent field. }:
\begin{align}
\psi&=\psi+c^{-1} \chi+c^{-2}\phi+\ldots \, ,\nonumber\\
C_i&= C_i+c^{-1} B_i+c^{-2}A_i+\ldots \, ,\label{ans}\\
\gamma_{ij}&= \gamma_{ij}+c^{-1}\alpha_{ij}+c^{-2}\beta_{ij}+\ldots \, .\nonumber
\end{align}
Furthermore
\begin{equation}
C_{ij}=\partial_iC_j-\partial_jC_i \, ,\qquad A_{ij}=\partial_iA_j-\partial_jA_i\, ,\qquad B_{ij}=\partial_iB_j-\partial_jB_i \, .
\end{equation}
As before we split the Lagrangians into a linear part and a source part, and additionally we organize the source part by the number of time derivatives:
\begin{align}
\os{\bar\call}{1}&=\sqrt{\g}\left(\os{\bar L}{1}_\mathrm{lin}+\os{\bar L}{1}_{1\mathrm{td}}\right)\, ,\label{LtotNLO}\\
\os{\bar\call}{2}&=\sqrt{\g}\left(\os{\bar L}{2}_\mathrm{lin}+\os{\bar L}{2}_{0\mathrm{td}}+\os{\bar L}{2}_{1\mathrm{td}}+\os{\bar L}{2}_{2\mathrm{td}}\right)\, .\label{LtotNNLO}
\end{align}  
Although the linear part is the same at all orders and was already computed above in \eqref{linLag} we will add it here as well, to provide a complete reference.

\subsubsection{NLO}\label{nlosec}
From the split of the relativistic Lagrangian (\ref{totlag}, \ref{L012}) it follows that
\begin{equation}
\os{\bar\call}{1}_{1\mathrm{td}}=\os{\bar{\call}}{0}_1 \, .
\end{equation}
An explicit calculation then yields
\begin{align}
\os{\bar L}{1}_\mathrm{lin}=&\,-\partial_i\psi\partial^i\chi+\frac{\chi}{2}e^{2\psi}C_{ij}C^{ij}+\frac{1}{2}e^{2\psi}C^{ij} B_{ij}-\bar{\a}^{ij}(R_{ij}+\frac{e^{2\psi}}{2}C_{ik}C_{j}{}^k-\frac{1}{2}\partial_i\psi\partial_j\psi)\, ,\nonumber\\
\os{\bar L}{1}_{1\mathrm{td}}=&\,\Gamma^{ij\,kl}D_iC_j\dot \g_{kl}+(2\partial^j\psi-e^{2\psi}C^{ij}C_i)\dot C_j +C^i\partial_i \psi\dot \psi  \, .\label{nlo}
\end{align}

\subsubsection{NNLO}\label{nnlosec}
At the next order the split of the relativistic Lagrangian (\ref{totlag}, \ref{L012}) implies
\begin{equation}
\os{\bar\call}{2}_{0\mathrm{td}}=\os{\bar\call}{2}_{0}-\os{\bar\call}{2}_{\mathrm{lin}}\, ,\qquad \os{\bar\call}{2}_{1\mathrm{td}}=\os{\bar\call}{1}_{1}\, ,\qquad \os{\bar\call}{2}_{2\mathrm{td}}=\os{\bar\call}{0}_{2} \, .
\end{equation}
After an explicit calculation one finds
	\begin{align}
	\os{\bar L}{2}_\mathrm{lin}=&-\partial_i\psi\partial^i\phi+\frac{\phi}{2}e^{2\psi}C_{ij}C^{ij}+\frac{1}{2}e^{2\psi}C^{ij}A_{ij}-\bar{\b}^{ij}(R_{ij}+\frac{e^{2\psi}}{2}C_{ik}C_{j}{}^k-\frac{1}{2}\partial_i\psi\partial_j\psi)\, ,\nonumber\\
	\os{\bar L}{2}_{0\mathrm{td}}=&-\frac{1}{4}\bar\a^{kl}\bar \a_{kl}(R-\frac{1}{2}\partial_i\psi\partial^i\psi+\frac{e^{2\psi}}{4}C_{ij}C^{ij})+\frac{1}{4}\Gamma^{ij\,kl\,mn}D_{i}\bar{\a}_{kl}D_{j}\bar{\a}_{mn}-\frac{1}{2}\partial_i\chi\partial^i\chi\nonumber\\
	&+\bar{\a}^{ik}\bar{\a}_k{}^j(R_{ij}-\frac{1}{2}\partial_i\psi\partial_j\psi+\frac{e^{2\psi}}{2}C_i{}^lC_{jl})+\bar{\a}^{ik}\bar{\a}^{jl}\frac{e^{2\psi}}{4}C_{ij}C_{kl}+\frac{1}{2}\chi^2e^{2\psi}C_{ij}C^{ij}\nonumber\\
	&+\bar{\a}^{ij}(\partial_i\psi\partial_j\chi-\chi e^{2\psi}C_{ik}C_j{}^k-e^{2\psi}B_{ik}C_j{}^k)+\frac{1}{4}e^{2\psi}B^{ij} B_{ij}+\chi e^{2\psi}C^{ij} B_{ij}\, ,\nonumber\\
	\os{\bar L}{2}_{1\mathrm{td}}=&  \,\, C_i \Big(  \dot{\gamma}  D_j \bar{\a}^{ij} + \dot{\gamma}_{jk} ( \frac{1}{2}D^i \bar{\a}^{jk} -  D^k \bar{\a}^{ij} )  -\dot{\psi}\bar\a^{ij} \partial_j \psi  + \dot{\psi} \partial^i\chi  + \dot{\chi} \partial^i\psi \Big)     \nonumber \\
	& - \dot{\gamma} \Big(D^i B_i- \bar{\a}^{ij}D_i C_j\Big)  + \dot{\gamma}^{ij} \Big(D_i B_j -\bar{\a}_{ik} D^k C_j - \bar{\a}_{jk} D_i C^k +\bar{\a}_{ij} D_k C^k  \Big)   \nonumber \\
	&+ B_i (\dot{\psi}\partial^i \psi  -e^{2\psi} \dot{C}_j C^{ij}  ) + 2\dot{C}_i ( \partial^i \chi- \bar\a^{ij} \partial_j \psi) + \dot{\bar{\a}}^{ij} D_i C_j - \dot{\bar\a} D^i C_i + 2 \dot{B}_i  \partial^i\psi   \nonumber  \\
	&+C_i e^{2\psi}\left[\dot{C}_j \Big(-2 \chi C^{ij} - B^{ij} + C_{kl} (\gamma^{ik} \bar\a^{jl} - \gamma^{jk}\bar\a^{il})\Big)   - \dot{B}_j C^{ij} \right] \nonumber \,,\\
	\os{\bar L}{2}_{2\mathrm{td}}=&-\Gamma^{ij\,kl}C_i\dot C_j\dot \gamma_{kl}+e^{-2\psi}\dot \gamma \dot \psi -2C^i \dot C_i\dot\psi+\frac{1}{4}(e^{-2\psi}\Gamma^{kl\,mn} +\Gamma^{ij\,kl\,mn}C_iC_j)\dot \gamma_{kl}\dot \gamma_{mn}\nonumber\\
	&-\frac{1}{2}e^{2\psi} \Gamma^{ij\,kl}C_iC_j\dot C_k\dot C_l-\frac{1}{2}( 3e^{-2\psi}+C_i C^i) \dot{\psi}^2 	\, .\label{nnlo}
	\end{align}

Previously only the leading order of an expansion of GR in $1/c$ including odd powers was computed \cite{Ergen:2020yop}, the results in sections \ref{nlosec} and \ref{nnlosec} are new. They show the computational advantage of the 3+1 formulation since the already quite involved expressions we obtained here would become quite a bit more involved to derive in their fully covariant Newton-Cartan form.

\section{The $1/c^2$ expansion in KS formalism}\label{evsec}

In this section we review how the $1/c$ expansion contains the $1/c^2$ expansion as a self-consistent sub-theory. In addition we spell out how the $1/c^2$ expansion up to order $c^{-2n}$ can be obtained from the $1/c$ expansion up to order $c^{-n}$ by a reshuffling of the terms. We use this to compute up to order $c^{-4}$ and compare to the results in the literature: in section \ref{secevnlo} we rederive the results of \cite{VandenBleeken:2017rij, Hansen:2020pqs} while in section \ref{secevnnlo} we show how a further truncation of our result reproduces the 1PN order of the PN expansion.

\subsection{Structure of the expansion}
\subsubsection{The even power ansatz and truncation}
The explicit form of the relativistic metric \eqref{metex} reveals that $c$ naturally appears with an odd power. This can be circumvented by defining
\begin{equation}
C_i=c^{-1}B_i\,,\label{evredef}
\end{equation} so that the metric takes the form 
\begin{equation}
ds^2=-c^{2}e^{\psi}(dt+c^{-2}B_i dx^i)^2+e^{-\psi}\g_{ij}dx^i dx^j \, .\label{metev}
\end{equation}
It is important to stress that this is much more than a simple redefinition when combined with the assumption, crucial for an expansion in $1/c$, that $C_i$ is analytic in $1/c$. Then \eqref{evredef} implies that $B_i$ is analytic, but more importantly also that $C_i$ is 'subleading'. I.e. the relation \eqref{evredef} should be interpreted as the non-trivial assumption that $\lim_{c\rightarrow\infty} C_i=0$.

Inserting the ansatz \eqref{evredef} into the KS action in the form (\ref{totlag}, \ref{L012}) one gets
\begin{equation}
\call^\mathrm{e}=\sqrt{\g}\left(L_0^\mathrm{e}+c^{-2}L_2^\mathrm{e}+c^{-4}L_4^\mathrm{e}+c^{-6}L_6^\mathrm{e}\right)\, ,\label{evLag}
\end{equation}
where now
\begin{align}
L_0^\mathrm{e}=&\,R-\frac{1}{2}\partial_i\psi\partial^i\psi \, ,\nonumber\\
L_2^\mathrm{e}=&\,\frac{1}{4}e^{2\psi}B_{ij}B^{ij}+\Gu^{ij\,kl}D_iB_j\dot \g_{kl}+2\partial^j\psi\dot B_j+B^i\partial_i \psi\dot \psi \nonumber\\
&+e^{-2\psi}\left(\frac{1}{4}\Gu^{kl\,mn}\dot \g_{kl}\dot \g_{mn}+\dot \g \dot \psi-\frac{3}{2} \dot{\psi}^2\right) \, ,\label{L0246}\\
L_4^\mathrm{e}=&-e^{2\psi}B^{ij}B_i\dot B_j-\Gu^{ij\,kl}B_i\dot B_j\dot \g_{kl} -2B^i \dot B_i\dot\psi\nonumber\\
&+\frac{1}{4}\Gu^{ij\,kl\,mn}B_iB_j\dot \g_{kl}\dot \g_{mn}-\frac{1}{2}B_i B^i \dot{\psi}^2 \, ,\nonumber\\
L_6^\mathrm{e}=&-\frac{1}{2}e^{2\psi}\Gu^{ij\,kl}B_iB_j\dot B_k\dot B_l \, .\nonumber
\end{align}
In summary, after the redefinition \eqref{evredef} both the metric \eqref{metev} and Lagrangian \eqref{evLag} contain only even powers of $c$. This in turn implies that if we assume the fields to be analytic in $1/c^2$, i.e.
\begin{equation}
\psi=\sum_{n=0}^\infty \os{\psi}{2n} c^{-2n} \, ,\qquad B_i=\sum_{n=0}^\infty \os{B}{2n}_i c^{-2n} \, ,\qquad \gamma_{ij}=\sum_{n=0}^\infty \os{\g}{2n}_{ij} c^{-2n}\, , \label{evexpansans}
\end{equation}
then we get a consistent expansion, that we'll refer to as the $1/c^2$ expansion. Furthermore this expansion is, via \eqref{evredef}, nothing but a truncation of the $1/c$ expansion:
\begin{equation}
(\mathrm{ev. trunc.})\,\qquad\os{\psi}{2m+1}\!\!\!=0,\quad \os{C}{2m}_i=0,\quad \os{C}{2m+1}\!\!\!{}_i=\os{B}{2m}_i,\quad\os{\g}{2m+1}\!\!\!{}_{ij}=0\qquad\forall m\geq 0  \, .\label{trunc}
\end{equation}
In particular, for the expansion coefficients of the action this implies
\begin{equation}
\os{S}{2n}^\mathrm{e}=\left.\os{S}{2n}\right|_{\mathrm{ev. trunc.}}\qquad 0=\!\!\!\os{S}{2n+1}\Big|_{\mathrm{ev. trunc.}}\, .
\end{equation}
At the level of the equations of motion observe that
\begin{eqnarray}
\os{\cale}{2n}\!{}_\psi^\mathrm{e}=\frac{\delta\!\! \os{S}{2n+2k}\!\!\!\!{}^\mathrm{e}}{\delta \os{\psi}{2k}}=\frac{\delta\Big(\os{S}{2n+2k}\Big|_{\mathrm{ev. trunc.}}\Big)}{\delta \os{\psi}{2k}}=\left.\frac{\delta\os{S}{2n+2k}}{\delta \os{\psi}{2k}}\right|_{\mathrm{ev. trunc.}}=\left.\os{\cale}{2n}\!{}_\psi\right|_{\mathrm{ev. trunc.}}\, ,\label{trunceom1}
\end{eqnarray}
The same observation holds for $\os{\cale}{2n}\!{}_\mathrm{e}^{ij}$,
\begin{eqnarray}
\os{\cale}{2n}{}_\mathrm{e}^{ij}=\frac{\delta\!\! \os{S}{2n+2k}\!\!\!\!{}^\mathrm{e}}{\delta \os{\gamma}{2k}\!{}_{ij}}=\left.\os{\cale}{2n}\!{}^{ij}\right|_{\mathrm{ev. trunc.}} \, ,
\end{eqnarray}
but gets a little twist in the case of $\os{\cale}{2n}\!{}_\mathrm{e}^{i}$:
\begin{eqnarray}
\os{\cale}{2n}{}_\mathrm{e}^{i}=\frac{\delta\!\! \os{S}{2n+2k}\!\!\!\!{}^\mathrm{e}}{\delta \os{B}{2k}{}_i}=\frac{\delta\Big(\os{S}{2n+2k}\Big|_{\mathrm{ev. trunc.}}\Big)}{\delta \os{B}{2k}_i}=\left.\frac{\delta\os{S}{2n+2k}}{\delta \os{C}{2k+1}\!\!\!{}_i}\right|_{\mathrm{ev. trunc.}}=\left.\os{\cale}{2n-1}^i\right|_{\mathrm{ev. trunc.}}\, .\label{trunceom2}
\end{eqnarray}
Finally note that 
\begin{equation}
0=\left.\os{\cale}{2n+1}\!\!\!{}_\psi\right|_{\mathrm{ev. trunc.}}\qquad  0=\left.\os{\cale}{2n+1}\!\!\!{}^{ij}\right|_{\mathrm{ev. trunc.}}\qquad 0=\left.\os{\cale}{2n}^i\right|_{\mathrm{ev. trunc.}} \, .
\end{equation}
This follows from the fact that the action as expressed in terms of $\psi, \gamma_{ij}$ and $B_i$ contains only even powers, see (\ref{evLag}, \ref{L0246}), and the same is thus true for the equations of motion $\cale_\psi=\frac{\delta S}{\delta \psi}$ and $\cale^{ij}=\frac{\delta S}{\delta \gamma_{ij}}$ when expressed in terms of the fields $\psi, \gamma_{ij}$ and $B_i$. Note that via \eqref{evredef} $\cale^{i}=\frac{\delta S}{\delta C_{i}}=c\frac{\delta S}{\delta B_{i}}$ and so this equation will only contain {\it odd} powers when expressed in terms of $\psi, \gamma_{ij}$ and $B_i$.

The discussion above shows in detail that the truncation \eqref{trunc} from the $1/c$ expansion to the $1/c^2$ expansion is a consistent truncation. This means it can be performed at the level of the action and that variation of the truncated action will reproduce the truncated equations of motion. Equivalently it also shows that the even coefficients do not source the odd coefficients, in case all of those are set to zero (the reverse is not true). We should point out that our discussion of the even power truncation is restricted to the pure gravitational or vacuum sector. In the presence of non-trivial energy momentum one needs to perform a similar analysis of that sector as well.

\subsubsection{The leading order as the static sector of GR}
The leading order of the $1/c^2$ expansion can be obtained directly by truncating the leading order of the $1/c$ expansion, which we discussed in section \ref{LOsec}. This simply amounts to removing the field $\os{C}{0}_{i}=0$ while keeping $\os{\psi}{0}$, $\os{\gamma}{0}_{ij}$. From now on, as we did previously, we will simply denote $\os{\psi}{0}$, $\os{\gamma}{0}_{ij}$ as $\psi$ and $\gamma_{ij}$ and leave it up to context to determine if they are to be interpreted as the relativistic fields or simply the leading coefficients in their expansion. The truncated Lagrangian \eqref{LOlag} takes the form
\begin{equation}
\os{\call}{0}^\mathrm{e}=\sqrt{\gamma}\left(R-\frac{1}{2}\partial_i\psi\partial^i\psi\right) \, .\label{LOevlag}
\end{equation}
This Lagrangian coincides with the fully relativistic one (\ref{totlag}, \ref{L012}) where all time derivatives and the field $C_i$ have been put to zero. Comparing to the form of the relativistic metric \eqref{metex} we see that extrema $(\psi,\gamma_{ij})$ of \eqref{LOevlag} coincide with quasi-static solutions of GR, i.e. static solutions with time dependent integration constants. This identification of the leading order of the $1/c^2$ expansion with the static sector of GR was previously made in \cite{VandenBleeken:2019gqa}.

\subsubsection{The universal linear part and partial decoupling}\label{evlinsec}
Let us recall from section \ref{linsec} that the $n$'th order equations of motion $\os{E}{n}$ are equations of motion determining the $n$'th order fields $\os{\Phi}{n}$. They take the form $\mathbb{D}^2\os{\Phi}{n}=\os{\mathbb{S}}{n}$. Here $\mathbb{D}^2$ is a second order linear differential operator that is universal, i.e. the same at all orders $n\geq 1$. The source term $\os{\mathbb{S}}{n}$ is a function of the lower order fields $\os{\Phi}{m}$, $m< n$, and their time derivatives, only. 
 
Since, as we discussed in detail above, the $1/c^2$ expansion is a truncation of the $1/c$ expansion, this structure carries over to the equations of motion $\os{E}{2n}\!{}_\mathrm{e}$ in this expansion. But due to the redefinition \eqref{evredef} we will need to treat  $\os{E}{2n}^{i}\!\!\!{}_\mathrm{e}$ separately from $\os{E}{2n}^\mathrm{e}\!\!\!{}_\psi$ and $\os{E}{2n}^{ij}\!\!\!\!{}_\mathrm{e}{}$. Defining $\os{\varPhi}{2n}=(\os{\psi}{2n},\os{\gamma}{2n}_{ij})$ the equations of motion in the $1/c^2$ expansion will take the form
\begin{equation}
\os{E}{2n}_\mathrm{e}=\mathbb{D}^2_\mathrm{e}\os{\varPhi}{2n}-\os{\mathbb{S}}{2n}_\mathrm{e}\qquad \os{E}{2n}_\mathrm{e}^i=\cald^2_\mathrm{e}\os{B}{2n-2}^i-\os{\mathbb{S}}{2n-2}_\mathrm{e}^i\qquad n\geq 1 \, .
\end{equation}
By combining (\ref{trunceom1}-\ref{trunceom2}) with \eqref{DD2} one finds the explicit form of universal the linear operators\footnote{Note that here, for simplicity, we again present the result in the traceless gauge (hence the bars), see section \ref{tracesec}. The extra terms appearing without this choice of gauge can be found in appendix \ref{traceapp}.}:
\begin{align}
\overline{\mathbb{D}}^2_\mathrm{e}\os{\psi}{2n}=& \,D_i\partial^i\os{\psi}{2n}+\partial_i\psi D_j\os{\bar\g}{2n}^{ij} \, ,\nonumber\\
\overline{\mathbb{D}}^2_\mathrm{e}\os{\bar\g}{2n}^{ij}=&\,\frac{1}{2}D^2 \os{\bar\g}{2n}^{ij}+\partial^{(i}\psi\partial^{j)}\os{\psi}{2n}-\frac{1}{2}\g^{ij}\partial_k\psi\partial^k\os{\psi}{2n}\nonumber\\
&+(\os{\bar\g}{2n}^{ik}\g^{jl}+\os{\bar\g}{2n}^{jk}\g^{il}-\frac{1}{2}\g^{ij}\os{\bar\g}{2n}^{kl})(R_{kl}-\frac{1}{2}\partial_k\psi\partial_l\psi)  \label{DDeven}\\
&- \frac{1}{3}\os{\bar\g}{2n}^{ij} (R - \frac{1}{2}\partial_k \psi \partial^k \psi)\, ,\nonumber\\
\bar\cald^2_\mathrm{e}\os{B}{2n+2}\!\!{}^j=&\,-D_i(e^{2\psi}\os{B}{2n+2}\!\!{}^{ij}) \, .\nonumber
\end{align}
where $D^2\os{\bar\g}{2n}^{ij}$ is defined in \eqref{D2}.

Note now that apart from a great simplification with respect to \eqref{DD2}, the linear operator \eqref{DDeven} is also block diagonal, so that the field $\os{B}{2n}_i$ decouples from the fields $\os{\varPhi}{2n}=(\os{\psi}{2n},\os{\gamma}{2n}_{ij})$. The operator $\mathbb{D}^2$ is fully off-diagonal, i.e. it mixes all components of $\os{\Phi}{n}=(\os{\psi}{n},\os{C}{n}_i,\os{\gamma}{n}_{ij})$. This has as a consequence that, unless one diagonalizes this operator, one has to solve for all $(\os{\psi}{n},\os{C}{n}_i,\os{\gamma}{n}_{ij})$ simultaneously at a given order $n$ in the $1/c$ expansion. On the contrary, at order $2n$ in the $1/c^2$ expansion, one can solve for $\os{B}{2n}_i$ independently of $(\os{\psi}{2n},\os{\gamma}{2n}_{ij})$ and vice versa. Schematically this result can be written as
\begin{equation}
\left.\mathbb{D}^2\right|_{\mathrm{ev.trunc.}}=\begin{pmatrix}
\mathbb{D}^2_\mathrm{e}&0\\
0& \cald^2_\mathrm{e}
\end{pmatrix}\, .
\end{equation}

\subsubsection{The $1/c^2$ expansion as a shuffled $1/c$ expansion.} \label{shufflingalg}
The $1/c^2$ expansion is a truncation of the $1/c$ expansion such that if one were to compute the $1/c$ expansion up to order $c^{-2N}$ and put all odd power coefficients to zero then one recovers the $1/c^2$ expansion up to order $c^{-2N}$. This is however not the most efficient way to relate the $1/c$ expansion to the $1/c^2$ expansion. If we compare the expansion ansatze \eqref{evexpansans} and \eqref{expansans} then we see they are equivalent under the replacement $c\rightarrow c^2$ and
\begin{equation}
\os{\psi}{n}\rightarrow \os{\psi}{2n}\, ,\qquad \os{C}{n}_i\rightarrow \os{B}{2n}_i\, ,\qquad \os{\g}{n}_{ij}\rightarrow \os{\g}{2n}_{ij}\label{shuffle} \, .
\end{equation}
This rather straightforward observation implies that the $1/c^2$ expansion will contain exactly the same terms as the $1/c$ expansion with the replacement \eqref{shuffle}. The key difference the replacement \eqref{shuffle} makes is that it changes the order of the various terms. As we'll now discuss, this change of order has a little twist, leading to a shuffling of terms among orders. But keeping track of the order and the shuffling is not too hard. 

We start by associating a weight\footnote{The weight is the power of $c^{-1}$ with which this object appears. The only factors of $c$ appearing are those associated to the weights as listed in (\ref{w1}, \ref{w2}). Please be aware that this notion of weight as we use it in this section is unrelated to the weight as defined in \eqref{tweights}} to the relevant objects in each expansion:
\begin{eqnarray}
1/c\mbox{ expansion:}&&
[\os{\gamma}{n}_{ij}]=[\os{\psi}{n}]=[\os{C}{n}_i]=n \, ,\qquad [\partial_t]=1\, .\label{w1}\\
1/c^2\mbox{ expansion:}&&
[\os{\gamma}{2n}\!{}_{ij}]=[\os{\psi}{2n}]=2n \, ,\quad [\os{B}{2n}_i]=2n+1 \, , \quad [\partial_t]=1\, .\label{w2}
\end{eqnarray}
Let us now consider terms in the $1/c^2$ expansion and trace back their origin in the $1/c$ expansion through the replacement \eqref{shuffle}. Every term will be polynomial\footnote{To be precise: polynomial in the coefficients of weight $\geq 1$. The zero weight fields can appear non-polynomially but this is irrelevant for our counting argument and so we can ignore that.} in the expansion coefficients. We can group $\Psi=(\psi,\gamma)$ since they have the same weight. Since spatial derivatives, as well as indices, have weight zero we can ignore them. We should keep track of time derivatives, but it is not relevant on which coefficient they act, so we will simply indicate the number of time derivatives at the beginning of the expression. So in this schematic fashion a term in the $1/c^2$ expansion has the form
\begin{equation}
T \sim \partial_t^\lambda\os{\Psi}{2k_1}^{\n_1}\ldots \os{\Psi}{2k_p}^{\n_p}\os{B}{2l_1}^{\m_1}\ldots \os{B}{2l_q}^{\m_q} \, .
\end{equation}
Please note that $\nu_a$ and $\mu_a$, as well as $\lambda\leq 2$, indicate positive integer powers. Via \eqref{w2} it follows that the weight of this term is
\begin{equation}
[T]=\lambda+2\sum_{a=1}^p k_a\nu_a+\sum_{a=1}^{q}(2l_a+1)\mu_a \, .
\end{equation}
Viewed through the replacement \eqref{shuffle}, the term $T$ originated from a term $\tilde T$ in the $1/c$ expansion, with
\begin{equation}
\tilde T\sim \partial_t^\lambda\os{\Psi}{k_1}^{\n_1}\ldots \os{\Psi}{k_p}^{\n_p}\os{C}{l_1}^{\m_1}\ldots \os{C}{l_q}^{\m_q} \, ,
\end{equation}
which, via \eqref{w1}, has weight
\begin{equation}
[\tilde T]=\lambda+\sum_{a=1}^p k_a\nu_a+\sum_{a=1}^{q}l_a\mu_a \, .
\end{equation}
Let us denote the weight of $T$ as $[T]=2N$ and the total number of $B$ (or $C$) coefficients as $M=\sum_a\mu_a$. One then finds that
\begin{equation}
[\tilde T]=N+\frac{\lambda-M}{2}\label{ordrel} \, .
\end{equation}
First of all, as a sanity check, remark that $\lambda-M$ is always even, guaranteeing $[\tilde T]$ is integer as it should. This follows by inspection of \eqref{L0246} or \eqref{L012} and observing that the terms with an even number of time derivatives come with even powers of $B$, respectively $C$, while the terms with one time derivative come with an odd power of $B$, respectively $C$. This is a consequence of invariance under the time redefinition symmetry \eqref{tred}, as discussed at the end of section \ref{expltimesec}.

Taking this into account, together with the fact that $M$ is positive by definition, we can conclude
\begin{equation}
[\tilde T_{0\mathrm{td}}]\leq N \, ,\qquad [\tilde T_{1\mathrm{td}}]\leq N \, , \qquad [\tilde T_{\mathrm{2td},\,C}]\leq N \, , \qquad [\tilde T_{\mathrm{2td},\, \mathrm{no}\,C}]=N+1  \, , \label{orderconnect}
\end{equation}
where $\tilde T_{\sharp\mathrm{td}}$ denotes terms with $\sharp$ time derivatives and we split the case with two time derivatives in those that contain coefficients of $C$ and those that do not.  

Formula \eqref{ordrel} shows that the presence of the $\os{B}{2n}_i/ \os{C}{n}_i$ and time derivatives shuffles orders under the replacement \eqref{shuffle} rather than simply relating them by a factor of two. The upshot of our discussion is however \eqref{orderconnect} which states that all terms up to order $c^{-2N}$ in the $1/c^2$ expansion originate from a term up to order $c^{-N}$ in the $1/c$ expansion, the only exception being those terms containing two time derivatives and no coefficients $\os{B}{2k}_i$, which originate from terms with two time derivatives and no $\os{C}{k}_i$ at order $c^{-N-1}$ in the $1/c$ expansion. 

We can transform the above conclusion to the following algorithm, which we'll refer to as {\it shuffling}, to obtain the $1/c^2$ expansion up to a given order $c^{-2N}$.

\paragraph{Shuffling algorithm}{\it Compute all terms in the $1/c$ expansion up to order $c^{-N}$. Make the replacement \eqref{shuffle}. Collect the resulting terms order by order by using the rule \eqref{w2} to determine the order. Compute the $c^{-N-1}$'th order in the $1/c$ expansion of the two time-derivative terms without $C_i$ factors in \eqref{L012}. Make the replacement \eqref{shuffle}. Add the result to the collection of terms at order $c^{-2N}$.} 

\vspace{0.3cm}

Note that under this procedure some terms will get a weight greater than $2N$, these can be discarded as they do not appear in the $1/c^2$ expansion up to order $c^{-2N}$. For this reason the $1/c^2$ expansion up to order $c^{-2N}$ is simpler than the $1/c$ expansion up to order $c^{-N}$. The shuffling algorithm can be applied to the Lagrangian as well as the equations of motion.

Let us compare the three ways to compute the $1/c^2$ expansion:
\begin{itemize}
	\item Direct approach\\
	Insert the ansatz \eqref{evexpansans} into the relativistic Lagrangian in the form (\ref{evLag}, \ref{L0246}). Expand up to order $c^{-2N}$.
	\item Shuffling\\
	Obtain the $1/c$ expansion up to order $c^{-N}$, use the shuffling algorithm outlined above to get the $1/c^2$ expansion up to order $c^{-2N}$.
	\item Truncation\\
	Obtain the $1/c$ expansion up to order $c^{-2N}$, put all odd power coefficients to zero to get the $1/c^2$ expansion up to order $c^{-2N}$.
\end{itemize}

In the absence of any results on the $1/c$ expansion the direct approach is the most efficient, since the $1/c^2$ expansion contains less terms than the $1/c$ expansion, and so it is easier to directly expand in $1/c^2$ up to order $c^{-2N}$ than it is to first expand in $1/c$ up to order $c^{-N}$ and then to shuffle\footnote{Note that just like the $1/c^2$ expansion, also the PN expansion is a truncation of the $1/c$ expansion. Similar to the $1/c^2$ expansion, a direct computation in the PN expansion will always be more efficient computationally than one via the $1/c$ expansion combined with shuffling and truncation.}. In case there is a result of the $1/c$ expansion up to some order $c^{-M}$ available, then the most efficient way to obtain the $1/c^2$ expansion up to order $c^{-2\lfloor \frac{M}{2}\rfloor}$ is of course to simply truncate. But shuffling is more powerful in that case, since it directly allows to reproduce the $1/c^2$ expansion all the way up to order $c^{-2(M-1)}$, and with the relatively little extra work of expanding the two time-derivative, no $C_i$, part of the Lagrangian \eqref{L012} to order $c^{-M-1}$ it provides the $1/c^2$ expansion up to order $c^{-2M}$.

We will use this in practice in section \ref{evexplsec}. Since we computed the $1/c$ expansion of the Lagrangian up to order $c^{-2}$ in section \ref{expsec} we can use the shuffling algorithm to easily find the $1/c^2$ expansion of the Lagrangian up to order $c^{-4}$, which goes beyond earlier results in the literature. At the next to leading order, i.e. order $c^{-2}$, one can explicitly see that the result obtained by shuffling matches with that obtained by truncation, as well as with results in the literature obtained by the direct approach.

\subsection{Explicit expansion up to NNLO}\label{evexplsec}
We will now perform the $1/c^2$ expansion explicitly up to NNLO, i.e. order $c^{-4}$. For notational convenience we indicate the various expansion coefficients with different symbols rather than with their superscript, as we did for the $1/c$ expansion. Furthermore we continue to use the same (abuse of) notation introduced in section \ref{LOsec}, indicating the leading order fields with the same symbol as the full $c$ dependent field. More precisely our expansion ansatz is
\begin{align}
\psi&=\psi+c^{-2}\phi+c^{-4}\tau+\ldots \, ,\nonumber\\
B_i&= B_i+c^{-2} Z_i+\ldots \, ,\label{evans}\\
\gamma_{ij}&=\gamma_{ij}+c^{-2}\beta_{ij}+c^{-4}\epsilon_{ij}+\ldots \, .\nonumber
\end{align}
With this notation the replacement \eqref{shuffle}, used in the shuffling algorithm, becomes
\begin{equation}
\chi\rightarrow \phi \, ,\quad \phi\rightarrow \tau \, , \quad C_i\rightarrow B_i \, ,\quad A_i\rightarrow Z_i \, ,\quad \alpha_{ij}\rightarrow\beta_{ij}\, ,\quad \beta_{ij}\rightarrow \epsilon_{ij} \, . \label{shuffling}
\end{equation}

We present the results for the expansion of the Lagrangian \eqref{evLag} as
\begin{equation}
\call^\mathrm{e}=\sqrt{\g}(\os{L}{0}^\mathrm{e}+c^{-2}\os{L}{2}^\mathrm{e}+c^{-4}\os{L}{4}^\mathrm{e}+\calo(c^{-6})) \, .
\end{equation}
Note that we will present these results in the traceless gauge, see section \ref{tracesec}, which we indicate by putting a bar on the relevant expressions. The extra terms appearing outside this gauge can be found in appendix \ref{traceapp}.
 
\subsubsection{NLO}\label{secevnlo}
The next to leading order, i.e. order $c^{-2}$, of the Lagrangian in the $1/c^2$ expansion can be most easily obtained by simply truncating $\os{\bar L}{2}$ as obtained in the $1/c$ expansion, see (\ref{LtotNNLO}, \ref{nnlo}). Alternatively one can also find it by performing the shuffling algorithm of section \ref{shufflingalg} to $\os{\bar L}{1}$, which is given in (\ref{LtotNLO}, \ref{nlo}). Both methods give the same result, which is
\begin{align}
\os{\bar L}{2}^\mathrm{e}=&\,\frac{1}{4}e^{2\psi}B_{ij}B^{ij}-\partial_i\psi\partial^i\phi-\bar{\b}^{ij}(R_{ij}-\frac{1}{2}\partial_i\psi\partial_j\psi)\nonumber\\
&+\Gamma^{ij\,kl}D_iB_j\dot \g_{kl}+2\partial^j\psi\dot B_j +B^i\partial_i \psi\dot \psi\label{L2evlag}\\
&+e^{-2\psi}\left(\frac{1}{4}\Gu^{kl\,mn}\dot \g_{kl}\dot \g_{mn}+\dot \g \dot \psi-\frac{3}{2} \dot{\psi}^2\right)\nonumber \, .
\end{align}
The $1/c^2$ expansion up to this order was first discussed at the level of the equations of motion in \cite{VandenBleeken:2017rij} and later at the level of the action in \cite{Hansen:2019pkl, Hansen:2020pqs}. The Lagrangian $\call_{\mathrm{NRG}}$ as given in (3.29) in \cite{Hansen:2020pqs} equals our $\os{\call}{2}^\mathrm{e}$ up to (irrelevant) total derivatives:
\begin{equation}
\call_{\mathrm{NRG}}=\os{\call}{2}^\mathrm{e}+\sqrt{\g}D_i\left(\phi \partial^i\psi-\Gamma^{ijkl}B_j(\dot \gamma_{kl} - \gamma_{kl}\,\dot{\psi})+2B^i\dot{\psi}\right)-2\partial_t(\sqrt{\g}B^i\partial_i\psi)\label{NRGrel}
\end{equation}
Here $\os{\call}{2}^\mathrm{e}$ is the Lagrangian density outside the traceless gauge, which via \eqref{superimportanttrace} is related to \eqref{L2evlag} as
\begin{equation}
\os{\call}{2}^\mathrm{e}=\sqrt{\g}\left(\os{\bar L}{2}^\mathrm{e}\big|_{\bar{\beta}_{ij}\rightarrow\beta_{ij}}+\frac{\beta}{2}(R-\frac{1}{2}\partial_i\psi\partial^i\psi)\right)\,.
\end{equation}
One can verify the equality \eqref{NRGrel} via the identification of our variables with those of \cite{Hansen:2020pqs}:
\begin{align}
\tau_\mu &= e ^{\psi/2}\ \delta_\mu^t  \,, \nonumber \\
h_{\mu\nu} &= e^{-\psi}  \  \delta_\mu^i \delta_\nu^j\ \gamma_{ij}\,,\nonumber \\
m_{\m} &=  e^{\psi/2} \left( \frac{1}{2}\delta_\mu^t \phi + \delta_\mu^i B_i \right)\,, \label{mapweyl} \\
\Phi_{\mu\nu} &=  e^{-\psi} \ \delta_\mu^i\delta_\nu^j\ \left( \beta_{ij} - \phi \gamma_{ij}\right)\,.\nonumber
\end{align}
One of the interesting insights provided in \cite{Hansen:2020pqs}, is that apart from obtaining the Lagrangian by expansion of the Einstein Hilbert action, it can also be constructed purely from symmetry considerations. Let us thus shortly comment on the relation between the symmetries as discussed in \cite{VandenBleeken:2017rij, Hansen:2020pqs} and their shape in the KS formalism.

The gauge parameters $\Lambda$ and $\xi^\mu$ in the transformations (\ref{tred}, \ref{texdsd}) are a priori themselves functions of $c^{-1}$ and should thus be expanded. Compatibility with the expansion ansatz \eqref{evans} requires\footnote{As previously we indicate the leading order of the vectorfield $\xi^i$ with the same symbol as the $c$ depedent vector field.}
\begin{align}
\Lambda &= c f + c^{-1} \lambda + \mathcal{O}(c^{-3})\,, \qquad \partial_i f=0 \, ,\label{lambtransf} \\
\xi^i &= \xi^i + c^{-2} \zeta^i+\calo(c^{-4})\,.
\end{align}
Ignoring $\delta_f$, we have three type of transformations of the fields. We can start with the leading order diffeomorphisms $\xi^i$:
\begin{align}
\delta_\xi \psi &=L_\xi\psi\,,&\quad \delta_\xi \gamma_{ij}&=L_\xi\gamma_{ij}\,,\quad \delta_\xi B_i=L_\xi B_i+e^{-2\psi}\gamma_{ij}\partial_t\xi^j\,,\\
\delta_\xi\phi&=L_\xi\phi+2B_i\partial_t\xi^i\,,& \delta_\xi \beta_{ij}&=L_\xi\beta_{ij}+2(B_k\gamma_{ij}+B_{(i}\gamma_{j)k})\partial_t\xi^k \, . &
\end{align}
These correspond to spatial diffeomorphisms also present in \cite{Hansen:2020pqs}, but note that the extra time derivative terms in the transformation of the subleading fields correspond to an additional Milne boost in the language of \cite{VandenBleeken:2017rij, Hansen:2020pqs}. The precise boost parameter in the conventions of \cite{Hansen:2020pqs} (HHO) is
\begin{equation}
\lambda_{\m}^{\mathrm{HHO}}=\delta_\m^i  e ^{-3\psi/2} \gamma_{ij} \partial_t\xi^j\,.
\end{equation}

For the $\lambda$ transformation \eqref{lambtransf} one finds
\begin{equation}
\delta_\lambda \psi=0\,,\,\,\,\delta_\lambda \gamma_{ij}=0\,,\,\,\, \delta_\lambda B_i=\partial_i\lambda\,,\,\,\,\delta_\lambda\phi=\lambda\dot{\psi}+2\dot{\lambda}\,,\,\,\,\delta_\lambda\gamma_{ij}=\lambda\dot{\gamma}_{ij}+2\dot\lambda\gamma_{ij} \, .\label{lambtrans}
\end{equation}
while the subleading diffeomorphisms parameterized by $\zeta^i$ lead to
\begin{eqnarray}
\delta_\zeta \psi=0\,,\quad \delta_\zeta \gamma_{ij}=0\,,\quad \delta_\zeta B_i=0\,,\quad
\delta_\zeta\phi=L_\zeta\psi\,,\quad \delta_\zeta \beta_{ij}=L_\zeta\gamma_{ij }\,.\label{zettrans}
\end{eqnarray}
The transformations (\ref{lambtrans}, \ref{zettrans}) are related to those of e.g. \cite{Hansen:2020pqs} via the field redefinition \eqref{mapweyl} and the following relation between the parameters
\begin{equation}
-\Lambda_{\mathrm{HHO}} v^\mu + h^{\mu\nu} \zeta_\nu^\mathrm{HHO}=\l \delta^\mu_t + e^{-\psi} \zeta^i  \delta^\mu_i \, .
\end{equation}

\subsubsection{NNLO}\label{secevnnlo}
We now proceed to the next order, i.e. order $c^{-4}$. As far as we are aware the gravitational action has not been previously expanded to this order, keeping all even power potentials as we do. As we will discuss below, upon further truncation of most of these potentials our result reproduces the post-Newtonian expansion at 1PN order.
Given our computation of $\os{L}{1}$ and $\os{L}{2}$, in the $1/c$ expansion, see sections \ref{nlosec}, \ref{nnlosec}, it is a surprisingly short calculation to obtain $\os{L}{4}^\mathrm{e}$ by the shuffling algorithm of section \ref{shufflingalg}. The result is:
\begin{align}
\!\!\!\os{\bar L}{4}^\mathrm{e}=&\,\,\frac{\phi}{2}e^{2\psi}B_{ij}B^{ij}+\frac{1}{2}e^{2\psi}B^{ij} Z_{ij}-\bar{\b}^{ij}\frac{e^{2\psi}}{2}B_{ik}B_{j}{}^k-e^{2\psi}B^{ij}B_i\dot B_j-\partial_i\psi\partial^i\tau\nonumber\\
&-\bar{\e}^{ij}(R_{ij}-\frac{1}{2}\partial_i\psi\partial_j\psi)-\frac{1}{4}\bar\b^{kl}\bar \b_{kl}(R-\frac{1}{2}\partial_i\psi\partial^i\psi)+\frac{1}{4}\Gamma^{ij\,kl\,mn}D_{i}\bar{\b}_{kl}D_{j}\bar{\b}_{mn}\nonumber\\
&-\frac{1}{2}\partial_i\phi\partial^i\phi+\bar{\b}^{ik}\bar{\b}_k{}^j(R_{ij}-\frac{1}{2}\partial_i\psi\partial_j\psi)+\bar{\b}^{ij}\partial_i\psi\partial_j\phi- \dot{\gamma} \Big(D^i Z_i- \bar{\b}^{ij}D_i B_j\Big)\nonumber\\
&+B_i \Big(  \dot{\gamma}  D_j \bar{\b}^{ij} + \dot{\gamma}_{jk} ( \frac{1}{2}D^i \bar{\b}^{jk} -  D^k \bar{\b}^{ij} )  -\dot{\psi}\bar\b^{ij} \partial_j \psi  + \dot{\psi} \partial^i\phi  + \dot{\phi} \partial^i\psi \Big)+ \dot{\bar{\b}}^{ij} D_i B_j\nonumber\\
& + \dot{\gamma}^{ij} \Big(D_i Z_j -\bar{\b}_{ik} D^k B_j - \bar{\b}_{jk} D_i B^k +\bar{\b}_{ij} D_k B^k  \Big)+ Z_i \dot{\psi}\partial^i \psi+ 2\dot{B}_i ( \partial^i \phi- \bar\b^{ij} \partial_j \psi)\nonumber\\
&- \dot{\bar\b} D^i B_i  + 2 \dot{Z}_i  \partial^i\psi-\Gamma^{ij\,kl}B_i\dot B_j\dot \gamma_{kl} -2B^i \dot B_i\dot\psi-\frac{1}{2}B_i B^i \dot{\psi}^2\label{evnnlo}\\
&+e^{-2\psi}\left( \frac{1}{2}\Gamma^{kl\,mn}\dot \gamma_{kl}\dot {\bar\beta}_{mn}+\frac{\bar{\beta}^{ij}}{2}(\dot{\gamma}_{ij}\dot{\gamma}-\dot{\gamma}_i{}^k\dot{\gamma}_{kj})+\dot {\bar\beta} \dot \psi-\bar{\beta}^{ij}\dot{\gamma}_{ij}\dot{\psi}+\dot{\gamma}\dot{\phi}-3 \dot{\psi}\dot{\phi}\right)\nonumber\\
&-2\phi e^{-2\psi}\left( \frac{1}{4}\Gamma^{kl\,mn}\dot \gamma_{kl}\dot \gamma_{mn}+\dot \gamma \dot \psi-\frac{3}{2} \dot{\psi}^2\right) +\frac{1}{4}\Gamma^{ij\,kl\,mn}B_iB_j\dot \gamma_{kl}\dot \gamma_{mn}\, .\nonumber
\end{align}
As before we presented the Lagrangian in trace-fixed form, see section \ref{tracesec}. The additional terms present outside this gauge choice are given in appendix \ref{traceapp}. 

Let us now discuss how the above Lagrangian describes an extension of the post-Newtonian expansion up to 1PN order. The metric in the 1PN approximation, see e.g. \cite{poisson_will_2014}, is
\begin{align}
ds^2 =&-c^2(1-2 Uc^{-2}-2(\Psi-U^2)c^{-4})dt^2-8 U_i c^{-2}dx^i dt\nonumber\\
&+(1+2Uc^{-2})dx^idx^i+\calo(c^{-4})\, .\label{1PNmet}
\end{align}
The Ricci tensor of this metric is
\begin{align}
R_{tt}&=-\partial_i\partial^i U+\left(-\partial_i\partial^i\Psi+4U\partial_i\partial^iU-4\partial_i\dot U^i-3\ddot U\right)c^{-2}+\calo(c^{-4})\, ,\label{1PN1}\\
R_{ti}&=\left(2\partial_jU^{j}{}_i-2\partial_i\dot U\right)c^{-2}+\calo(c^{-4})\, ,   \qquad U_{ij} = \partial_i U_j - \partial_j U_i\, ,\\
R_{ij}&=-\partial_k\partial^k U \delta_{ij} c^{-2}+\calo(c^{-4})\, .\label{1PN3}
\end{align}
and the coefficients of the various powers of $c^{-2}$ provide the vacuum gravitational equations up to 1PN order. If we compare \eqref{1PNmet} to our relativistic metric \eqref{metev} and expansion ansatz \eqref{evans} one finds the identification
\begin{eqnarray}
\psi=0\,,\,\,\, \gamma_{ij}=\delta_{ij}\,,\,\,\, B_i=0\,,\,\,\,\phi=-2U\, ,\,\,\,\beta_{ij}=0\,,\,\,\, Z_i=4U_i\,,\,\,\, \tau=-2\Psi\, .\label{1PNus}
\end{eqnarray}
while the field $\epsilon_{ij}$ remains undetermined by the 1PN metric ansatz \eqref{1PNmet}.
  
One sees that most of the fields that a priori can be non-trivial in the $1/c^2$ expansion are assumed to be zero in the 1PN expansion. This is because, by definition, the Post-Newtonian expansion is a non-relativistic expansion around flat space, making it a weak gravity expansion as well. The $1/c^2$ expansion is however a non-relativistic expansion around an arbitrary quasi-static metric and includes certain non-linear, strong gravitational effects exactly.

Although it would still be quite an effort to derive the full equations of motion $\os{E}{4}$ from \eqref{evnnlo}, it becomes rather easy under the assumption \eqref{1PNus} where most fields are set to trivial values. From $\eqref{evnnlo}$ together with \eqref{LOevlag} and \eqref{L2evlag} we can find all equations of motion up to order $c^{-4}$, the result is\footnote{It is rather straightforward to verify that the trace terms do not contribute under the ansatz \eqref{1PNus}, so one can work with the trace fixed Lagrangians without loss of generality.}
\begin{align}
\os{E}{0}^\mathrm{e}_\psi&=0 \, ,\qquad\qquad \os{E}{0}_\mathrm{e}^{ij}=0 \, ,\qquad\qquad \os{E}{0}_\mathrm{e}^i=0 \, .\\
\os{E}{2}^\mathrm{e}_\psi&=\partial_i\partial^i\phi=-2\partial_i\partial^iU \, ,\quad\qquad \os{E}{2}_\mathrm{e}^i=0 \, ,\quad\qquad \os{E}{2}_\mathrm{e}^{ij}=0 \, .\\
\os{E}{4}^\mathrm{e}_\psi&=\partial_i\partial^i\tau-2\partial_i\dot Z^i+3\ddot\phi=-2\partial_i\partial^i\Psi-8\partial_i\dot{U^i}-6\ddot U \, ,\\
\os{E}{4}_\mathrm{e}^i&=-\partial_j Z^{ji}-2\partial^i\dot{\phi}=-4\partial_j U^{ji}+4\partial^i\dot{U} \, .
\end{align}
Here we see explicitly how the equations of motion in the $1/c^2$ expansion up to order $c^{-4}$ reproduce (a non-degenerate linear combination of) the 1PN equations (\ref{1PN1}-\ref{1PN3}). Note that we did not include the equation of motion $\os{E}{4}_\mathrm{e}^{ij}$, since it involves the field $\epsilon_{ij}$, which is of higher than 1PN order.

\section{Discussion}\label{discsec}
In this paper we revisited the $1/c$ expansion starting from the KS 3+1 formulation of GR, a lesser known dual version of the better known ADM decomposition. Although this 3+1 formulation renders space-time diffeomorphism invariance non-manifest, it preserves manifest spatial diffeomorphism invariance and keeps the variational principle intact. It has the advantage it makes the degrees of freedom more explicit, with the 4 dimensional Lorentzian metric being parameterized by a scalar $\psi$, a vector field $C_i$ and a metric $\gamma_{ij}$, with the last two carrying spatial indices only. These fields are assumed to be analytic in the inverse speed of light $1/c$ and the coefficients in their series expansion form the effective fields of the $1/c$ expansion. All effective fields up to order $c^{-4}$ are listed at the top of table \ref{fieldstable}. The 3+1 formulation makes the structure of the expansion more transparent and that allowed us to compute the effective Lagrangians to higher order than before.  We extended the computation of the Lagrangians from leading order \cite{Ergen:2020yop} to next to next to leading order in case of the $1/c$ expansion including odd terms, which is described by the fields listed in the second table from the left in table \ref{fieldstable}. In addition we made some all order observations as well, with a computation of the universal linear part of the expanded equations of motion and an all order gauge fixing of the trace of the metric coefficients. The 3+1 formulation also clarifies the relation between the $1/c$ and $1/c^2$ expansions, the latter being a consistent truncation to even powers of the former. This allowed us to formulate an algorithm to compute the $1/c^2$ expansion up to order $c^{-2N}$ from the $1/c$ expansion to order $c^{-N}$. Using this shuffling algorithm we obtained the Lagrangians describing the $1/c^2$ expansion to order $c^{-4}$. At order $c^{-2}$ this matches the earlier results of \cite{VandenBleeken:2017rij, Hansen:2020pqs}, at order $c^{-4}$ the result is new. A further truncation simplifies the $1/c^2$ expansion to the PN expansion -- in table \ref{fieldstable} one finds all potentials included in the $1/c^2$ expansion up to order $c^{-4}$ listed at the second right and the corresponding potentials in the PN expansion to 1PN order on the far right -- and in that case our result simplifies to the standard expressions \cite{poisson_will_2014}.

\begin{table}
	\begin{center}
		
		\begin{tabular}{|c|c|c|}
		 $\psi$ & $C_i$ & $\gamma_{ij}$\\
			\hline
			$\os{\psi}{0}$ &  $\os{C}{0}_i$ & $\os{\gamma}{0}_{ij}$\\
		$\os{\psi}{1}$ &  $\os{C}{1}_i$ & $\os{\gamma}{1}_{ij}$\\
			$\os{\psi}{2}$ &  $\os{C}{2}_i$ & $\os{\gamma}{2}_{ij}$\\
			$\os{\psi}{3}$ &  $\os{C}{3}_i$ & $\os{\gamma}{3}_{ij}$\\
			$\os{\psi}{4}$ &  $\os{C}{4}_i$ & $\os{\gamma}{4}_{ij}$\\
		\end{tabular} \\ \vspace{0.2cm} $1/c$ expansion\\ pedantic notation \\ \vspace{0.3cm} 
		\begin{tabular}{c}
		{\rule{0pt}{3ex}}\\
			$c^{0}$\\
			$c^{-1}$\\
			$c^{-2}$\\
			$c^{-3}$\\
			$c^{-4}$\\
		\end{tabular}
		\begin{tabular}{|c|c|c|}
			$\psi$ & $C_i$ & $\gamma_{ij}$\\
			\hline
		     	$ \psi$ &  $C_i$ & $\gamma_{ij}$\\
		        		$\chi$ &  $B_i$ & $\a_{ij}$\\
			 $\phi$ &  $A_i$ & $\b_{ij}$\\
			 	$\upsilon$ &  $Z_i$ & $\varpi_{ij}$\\
			 	$\tau$ &  $Y_i$ & $\epsilon_{ij}$\\
		\end{tabular} \qquad \begin{tabular}{|c|c|c|}
			$\psi$ & $C_i$ & $\gamma_{ij}$\\
			\hline
			$\psi$ &  $C_i$ & $\gamma_{ij}$\\
			$\chi$ &  $B_i$ & $\a_{ij}$\\
			$\phi$ &  $A_i$ & $\b_{ij}$\\
			&   & \\
			&   & \\
		\end{tabular}\qquad
		\begin{tabular}{|c|c|c|}
			$\psi$ & $C_i$ & $\gamma_{ij}$\\
			\hline
			$\psi$ &  $0$ & $\gamma_{ij}$\\
			$0$ &  $B_i$ & 0\\
			$\phi$ &  $0$ & $\b_{ij}$\\
			$0$ &  $Z_i$ & $0$\\
			$\tau$ &  0 & $\epsilon_{ij}$\\
		\end{tabular} \qquad \begin{tabular}{|c|c|c|}
			$\psi$ & $C_i$ & $\gamma_{ij}$\\
			\hline
			$0$ &  $0$ & $\delta_{ij}$\\
			$0$ &  $0$ & 0\\
			$-2U$ &  $0$ & $0$\\
			$0$ &  $4 U_i$ & $0$\\
			$-2\Psi$ & 0  & \\
		\end{tabular}
	\end{center}  \vspace{-0.3cm}
	
	\quad\qquad $1/c$ expansion\qquad\, $1/c$ expansion\qquad\!\!\! $1/c^2$ expansion\qquad\ \ \ PN expansion \\ 
	
	\vspace{-0.5cm}
	
	\qquad\quad\ \ alt notation\qquad\quad up to NNLO\qquad\ \ up to NNLO\qquad\quad\ \ up to 1PNO
	\caption{On the top all potentials/fields in the $1/c$ expansion up to order $c^{-4}$ are listed in a precise but somewhat pedantic notation. On the left of the second line the same potentials are listed in an alternative notation. Second to left are those potentials featuring in the $1/c$ expansion up to NNLO, while second to right are those potentials present in the truncation to the $1/c^2$ expansion up to NNLO. All the way on the right we list those potentials present in the PN expansion up to first Post-Newtonian order. On the far left the order at which these fields enter the $1/c$ expansion of their relativistic relativistic counterparts is indicated.}\label{fieldstable}
\end{table}

We believe that in the form introduced in this paper, the $1/c$ expansion is now finally ready to be applied. Rather than a push to ever increasing order, the priority in the near future should be to explore non-trivial solutions of the equations of motion, an interpretation of the physics they describe and to study if and how this might improve on results obtained by the PN expansion. That the $1/c$ expansion is not an empty theory is well established. Various example solutions are discussed in \cite{VandenBleeken:2017rij, Hansen:2020pqs, Ergen:2020yop} but these all take the form of expansions of exact solutions of GR. Although such examples are useful to gain intuition into some features of the expansion they are not teaching us anything inherently new. Of greater interest would be solutions to the $1/c$ expansion that provide approximations to solutions of GR in situations where no exact solution is known. Natural situations to think of would be strong gravitational systems such as Neutron stars or gravitational dynamics close to merger, where the PN expansion would be put to the limit and the extra potentials included in the $1/c$ expansion could play a crucial role. We hope the work in this paper will have paved some of the way for future work in this direction.

Let us shortly return to the ADM and KS formulations of general relativity. Our discussion in section \ref{dualsec} showed how both can be implemented in a unified way, where they appear as each others dual, in that the first is based on a preferred frame $e_i=\partial_i, u=N^{-1}(c^{-1}\partial_t-N^i\partial_i)$ while the second is base on the particular co-frame $e^i=dx^i, n=-M(cdt+C_idx^i)$. The frame becomes degenerate in the $c\rightarrow \infty$ limit, but remains generating in the $c\rightarrow 0$ limit, while the opposite is true for the co-frame. This implies the KS formalism is the natural 3+1 decomposition to use in the non-relativistic or galilean limit, while the ADM formalism is well suited for the ultra-relativistic or carrolian limit. This interpretation suggests a link between KS/ADM duality and Galilei/Carroll duality \cite{Duval:1990hj, Duval:2014uoa, Barducci:2018wuj, Figueroa-OFarrill:2022pus, Bergshoeff:2022qkx}. In the ADM, respectively KS, formalism the Lorentzian metric takes the form \eqref{metcase1}, respectively \eqref{metcase2} where the fields $(N,N^i,h_{ij})$  respectively $(M,C_i,h_{ij})$ are functions of both time $t$ and space, $x^i$. If one however assumes these fields to be time independent then both metrics \eqref{metcase1} and \eqref{metcase2} are stationary metrics. These two different forms of a stationary metric are known as the Zermelo and Randers (-Papapetrou) forms. The two forms can also be argued to be dual \cite{Gibbons:2008zi} and appear in carrolian and galilean fluid dynamics, see e.g. \cite{Ciambelli:2018xat, Petkou:2022bmz}. A better understanding of the relation between KS/ADM duality and Galilei/Carroll as well as Randers/Zermelo duality would be interesting on its own and might be applicable to the $1/c$ expansion or PN expansion as well, see e.g. \cite{Kol:2010ze}.  

Finally let us mention that in the last few years the study of nonrelativistic gravity per se -- i.e. independent of a relativistic counterpart -- has been very active, see e.g. \cite{Papageorgiou:2009zc, Bergshoeff:2016lwr, Aviles:2018jzw, Ozdemir:2019orp, Bergshoeff:2020fiz, Grumiller:2020elf, Gomis:2020wxp, Gallegos:2020egk, Concha:2020eam, Bergshoeff:2021bmc, Hartong:2021ekg, Ravera:2022buz, Concha:2022jdc}, motivated by quantum gravity, holography and string theory as well as condensed matter applications.  The $1/c^2$ expansion has provided some concrete examples of nonrelativistic gravity theories and provides a rather generic technique to obtain them. This has influenced advancement in the wider field of nonrelativistic gravity as well and we hope the same might be true for the results and insights provided in this paper, in particular as it is the first development of the $1/c$ expansion beyond leading order keeping all odd power coefficients.

\section*{Acknowledgements}
We thank  I. Aras, D. Demirhan, J. Figueroa-O'Farrill, G. Gibbons, P. Horvathy and N. Obers for discussions and/or correspondence, and the anonymous referee for a careful reading of the manuscript and valuable comments and suggestions. DVdB and ME were partially supported by the Boğaziçi University Research Fund under grant number 21BP2. ME is currently supported by TÜBİTAK under 2236-Co-Funded Brain Circulation Scheme2 (CoCirculation2) with project number 121C356. UZ is supported by TÜBİTAK - 2218 National Postdoctoral Research Fellowship Program with grant number 118C512. DVdB was also partially supported by the Bilim Akademisi through a BAGEP award.
		
\appendix

\section{Additional technical arguments}\label{techap}
\subsection{Expanded action and equations of motion}\label{acteomapp}
The relation between the expanded action\footnote{Since Lagrangians are ambiguous up to total derivatives we find it convenient to perform the discussion at the level of the action, but the discussion passes over to Lagrangians immediately.} and expanded equations of motion is independent of the particular theory under question, so we will keep it general. Apart from setting some definitions and notation the main result derived in this appendix that is of use in the main text is \eqref{exprels}.

Let $S[\Phi;\epsilon]$ be an action functional of some fields $\Phi$, that can have some explicit dependence on an expansion parameter $\epsilon$. In the case of interest to this paper $\Phi=(\psi,C_i,\gamma_{ij})$ and $\epsilon=1/c$. We'll denote the Euler-Lagrange equations associated to this action as $\cale[\Phi;\epsilon]=\frac{\delta S}{\delta \Phi}$. We assume the fields themselves to be analytic functions of $\epsilon$: $\Phi=\sum_{n=0}^\infty \os{\Phi}{n}\epsilon^n$. 

Let us then consider an expansion order by order in $\epsilon$, of the equations of motion and the action. We can write
\begin{align}
S[\Phi;\epsilon]&=\sum_{n=0}^\infty \os{S}{n}[\os{\Phi}{0},\ldots, \os{\Phi}{n}]\epsilon^n\label{actexp} \, ,\\
\cale[\Phi;\epsilon]&=\sum_{n=0}^\infty\os{\cale}{n}[\os{\Phi}{0},\ldots, \os{\Phi}{n}]\epsilon^n \, .
\end{align}
Note that if we expand the theory up to order $N$ then we would have $N$ equations of motion, but also $N$ actions that can be varied with respect to $N$ fields. At first there might appear a mismatch, but the key insight \cite{Hansen:2020pqs} is that there is a large degeneracy between the variations of the expanded actions. Indeed, as is shown below, one has the following relations:
\begin{equation}
\frac{\delta \os{S}{n+m}}{\delta \os{\Phi}{m}}=\frac{\delta \os{S}{n}}{\delta \os{\Phi}{0}}=\os{\cale}{n}\, .\label{exprels}
\end{equation}
To see this, first observe that:
\begin{equation}
\delta \Phi=\sum_{n=0}^\infty \delta\os{\Phi}{n}\epsilon^n\, ,
\end{equation}
which implies that
\begin{equation}
\frac{\delta S}{\delta \os{\Phi}{n}}=\frac{\delta S}{\delta \Phi}\frac{\delta\Phi}{\delta \os{\Phi}{n}}=\frac{\delta S}{\delta \Phi}\epsilon^n \, .\label{eq1}
\end{equation}
This equation has two consequences. First note that in the case $n=0$ it gives $\frac{\delta S}{\delta \Phi}=\frac{\delta S}{\delta \os{\Phi}{0}}$ so we get
\begin{equation}
\sum_{n=0}^\infty\os{\cale}{n}\epsilon^n=\cale=\frac{\delta S}{\delta \Phi}=\frac{\delta S}{\delta \os{\Phi}{0}}=\sum_{n=0}^\infty \frac{\delta \os{S}{n}}{\delta \os{\Phi}{0}}\epsilon^n \, ,
\end{equation}
and this implies
\begin{equation}
\frac{\delta \os{S}{n}}{\delta \os{\Phi}{0}}=\os{\cale}{n} \, .
\end{equation}
Secondly, it directly follows from \eqref{eq1} that
\begin{equation}
\frac{\delta S}{\delta \os{\Phi}{n+1}}=\frac{\delta S}{\delta \os{\Phi}{n}}\epsilon \, .
\end{equation}
Via \eqref{actexp} this becomes
\begin{equation}
\sum_{k=0}^\infty\frac{\delta\os{S}{k}}{\delta \os{\Phi}{n}}\epsilon^{k+1}=\sum_{k=0}^\infty\frac{\delta \os{S}{k}}{\delta \os{\Phi}{n+1}}\epsilon^k=\sum_{k=1}^\infty\frac{\delta \os{S}{k}}{\delta \os{\Phi}{n+1}}\epsilon^{k}=\sum_{k=0}^\infty\frac{\delta \os{S}{k+1}}{\delta \os{\Phi}{n+1}}\epsilon^{k+1} \, .
\end{equation}
Here we used that $\os{S}{0}[\os{\Phi}{0}]$, i.e. the leading order action is a functional of the leading order fields only, and that $n\geq 0$. Matching the coefficients then gives
\begin{equation}
\frac{\delta \os{S}{k+1}}{\delta \os{\Phi}{n+1}}=\frac{\delta \os{S}{k}}{\delta \os{\Phi}{n}} \, .
\end{equation}
Applying this relation $m$ times leads to \eqref{exprels}.

\subsection{Gauge fixing the traces: details}\label{techtraceapp}
In this appendix we discuss in detail how the traces $\os{\gamma}{n}=\os{\g}{0}^{ij}\os{\gamma}{n}_{ij}$, $n\geq 1$, can be put to zero by a choice of coordinate gauge and then show the equivalence \eqref{eqeoms}. We conclude with a summary of the result and provide a detailed description of the variational principle \eqref{gfvar}. 

\subsubsection{Gauge fixing}
To show how using the time-redefinition symmetry \eqref{tred} we can always remove the traces $\os{\g}{n}$, we start from the time-redefinition symmetry \eqref{tred}, which in expanded form gives
\begin{equation}
	\delta \os{\g}{n}_{ij}=\sum_{m=0}^{n-1}\left(2\os{\g}{n-m-1}\!\!\!\!\!\!{}_{ij}\ \partial_t \os{\Lambda}{m} +\partial_t\!\!\os{\g}{n-m-1}\!\!\!\!\!\!{}_{ij}\ \os{\Lambda}{m}\right) \,.
\end{equation}
The first observation is that $\os{\gamma}{0}_{ij}$ does not\footnote{Consistency of \eqref{tred} with our expansion ansatz requires a form $\Lambda(t,x)=c\, f (t)+\sum_{n=0}^\infty \os{\Lambda}{n}(t,x)c^{-n}$. Since $f$ is a function only of $t$ it cannot be used to set a field to zero and so it is irrelevant in the discussion here, so we can put it to zero without loss of generality.} transform, i.e. $\delta\os{\gamma}{0}_{ij}=0$, since the transformation \eqref{tred} of $\gamma_{ij}$ is suppressed by a factor $c^{-1}$. Using this we get
\begin{align*}
	\delta \os{\g}{n}=\delta(\os{\g}{0}^{ij}\os{\g}{n}_{ij})=\os{\g}{0}^{ij}\delta\os{\g}{n}_{ij}=&\,6\partial_t\!\!\os{\Lambda}{n-1}+\os{\g}{0}^{ij}\partial_t\os{\g}{0}_{ij}\os{\Lambda}{n-1}\\
	&+\sum_{m=0}^{n-2}\left(2\os{\g}{n-m-1}\partial_t \os{\Lambda}{m} +\os{\Lambda}{m}\os{\g}{0}^{ij}\partial_t\os{\g}{n-m-1}\!\!\!\!{}_{ij} \right) \, .
\end{align*}
From this it follows that we can iteratively set all $\os{\gamma}{n}$, $n\geq 1$, to zero. To see this explicitly let us define
\begin{equation}
	\os{\Lambda}{n}_\star=-\frac{e^{-\frac{1}{6}\os{\gamma}{0}}}{6}\int^t e^{\frac{1}{6}\os{\gamma}{0}}\os{\gamma}{n+1} dt' \, , 
\end{equation}
where $\os{\gamma}{0}=\det(\os{\gamma}{0}_{ij})$. As the first step then, observe that the gauge transformation with parameters $\os{\Lambda}{0}=\os{\Lambda}{0}_\star$ and $\os{\Lambda}{m}=0$ when $m\neq 0$, puts $\os{\gamma}{1}$ to zero. In a second step we can then put (the new) $\os{\gamma}{2}$ to zero by a transformation with parameters $\os{\Lambda}{1}=\os{\Lambda}{1}_\star$ and $\os{\Lambda}{m}=0$ when $m\neq 1$. The key point is that the transformation of the second step does not alter $\os{\gamma}{1}$ (which we previously set to zero), since $\os{\gamma}{1}$ does only transform with $\os{\Lambda}{0}$. Indeed, since $\os{\gamma}{n}$ only transforms under the $\os{\Lambda}{m}$ for which $m<n$, we see we can keep on repeating these steps in such a way that when we put $\os{\gamma}{n}$ to zero by the transformation with $\os{\Lambda}{n-1}=\os{\Lambda}{n-1}_\star$ and $\os{\Lambda}{m}=0$ when $m\neq n-1$, all the $\os{\gamma}{m}$ with $m<n$ (that we previously set to zero) will remain zero.

We thus conclude we can always redefine the time coordinate (while keeping the metric in the form \eqref{metex}) such that all $\os{\gamma}{n}$ are zero.

\subsubsection{Equivalence to gauge fixed action}
Here we provide the argument leading to the equivalence \eqref{eqeoms}. Since the $\os{\gamma}{n}$ are independent of $\os{\psi}{0}$ and $\os{C}{0}_i$ it immediately follows that\footnote{To ease notation we write simply ${\os{\gamma}{m}=0}$, rather than $\os{\gamma}{m}=0, \forall m\geq 1$.}
\begin{align}
	\left.\os{\cale}{n}_\psi\right|_{\os{\gamma}{m}=0}&=\left.\frac{\delta \os{S}{n}}{\delta\os{\psi}{0}}\right|_{\os{\gamma}{m}=0}=\frac{\delta \left( \os{S}{n}|_{\os{\gamma}{m}=0}\right)}{\delta\os{\psi}{0}}=\frac{\delta\os{\bar S}{n}}{\delta \os{\psi}{0}}=\os{\bar\cale}{n}_\psi \, ,\label{equiv1}\\
	\left.\os{\cale}{n}^i\right|_{\os{\gamma}{m}=0}&=\left.\frac{\delta \os{S}{n}}{\delta\os{C}{0}_i}\right|_{\os{\gamma}{m}=0}=\frac{\delta \left( \os{S}{n}|_{\os{\gamma}{m}=0}\right)}{\delta\os{C}{0}_i}=\frac{\delta\os{\bar S}{n}}{\delta \os{C}{0}_i}=\os{\bar\cale}{n}^i \, .\label{equiv2}
\end{align}
The same is not true for the variation with respect to $\os{\gamma}{0}_{ij}$. Since $\os{\gamma}{m}=\os{\gamma}{0}^{ij}\os{\gamma}{n}_{ij}$ it follows that
\begin{equation}
	\left.\frac{\delta \os{S}{n}}{\delta\os{\gamma}{0}_{ij}}\right|_{\os{\gamma}{m}=0}\neq\frac{\delta \left( \os{S}{n}|_{\os{\gamma}{m}=0}\right)}{\delta\os{\gamma}{0}_{ij}} \, .
\end{equation}
To understand more precisely how these two objects are related we split $\os{\gamma}{m}_{ij}$ in its trace and trace-less part, as defined in \eqref{tracesplit}, so that\footnote{In \eqref{actions} again we use some shorthand notation, where $\os{S}{n}[\os{\psi}{0}, \os{C}{0}_i, \os{\gamma}{0}_{ij};\os{\psi}{m}, \os{C}{m}_i,\os{\gamma}{m}_{ij}]$ should be understood as $\os{S}{n}[\os{\psi}{0}, \os{C}{0}_i, \os{\gamma}{0}_{ij};\os{\psi}{1},\ldots, \os{\psi}{n},  \os{C}{1}_i, \ldots , \os{C}{n}_i, \os{\gamma}{1}_{ij},\ldots, \os{\gamma}{n}_{ij}]$ and similarly for $\os{\tilde S}{n}$.} 
\begin{equation}
	\os{S}{n}[\os{\psi}{0}, \os{C}{0}_i, \os{\gamma}{0}_{ij};\os{\psi}{m}, \os{C}{m}_i,\os{\gamma}{m}_{ij}]=\os{\tilde S}{n}[\os{\psi}{0}, \os{C}{0}_i,\os{\gamma}{0}_{ij};\os{\psi}{m}, \os{C}{m}_i,\os{\bar\gamma}{m}_{ij},\os{\bar\gamma}{m}] \, .\label{actions}
\end{equation}
One can think of the replacement $(\os{\gamma}{0}_{ij};\os{\gamma}{m}_{ij})$ by $(\os{\gamma}{0}_{ij};\os{\bar\gamma}{m}_{ij},\os{\gamma}{m})$ as a change of variables. There is an important subtlety however, in that the new variables are redundant and not all independent, indeed the collection of variables $(\os{\gamma}{0}_{ij};\os{\bar\gamma}{m}_{ij},\os{\gamma}{m})$ satisfy the constraints
\begin{equation}
	\os{\g}{0}^{ij}\os{\bar\g}{m}_{ij}=0 \,.\label{constr}
\end{equation}
In practice this has the effect that while we can keep all $\os{\gamma}{m}$ fixed while varying $\os{\gamma}{0}_{ij}$, this is not the case for the $\os{\bar{\gamma}}{m}_{kl}$. We can however require them to vary in as minimal\footnote{Remark that $\eqref{constr}$ only constrains the trace of the variation of $\os{\bar{\gamma}}{m}_{ij}$, enforcing it to be non-zero. A priori one could consider adding a traceless part (in $kl$) to the right hand side of \eqref{subtle}. Excluding this, as we do, amounts to taking $\os{\bar\gamma}{m}_{ij}$ as independent as possible of $\os{\gamma}{0}_{kl}$.} a way as possible, while preserving \eqref{constr}. This leads to
\begin{equation}
	\frac{\partial \os{\bar\g}{m}_{kl}}{\partial \os{\g}{0}_{ij}}=\frac{1}{3}\os{\gamma}{0}_{kl}\os{\bar{\g}}{m}^{ij} \, .\label{subtle}
\end{equation}
In reverse, it is possible to keep $\os{\g}{m}$ and $\os{\g}{0}_{ij}$ fixed when varying the $\os{\bar\gamma}{m}_{ij}$, if we do so in a limited fashion. More precisely we only allow the $\os{\bar\gamma}{m}_{ij}$ to vary in such a way that this preserves \eqref{constr} while $\os{\g}{0}_{ij}$ is kept fixed. This is equivalent to
\begin{equation}
	\frac{\partial \os{\bar\g}{m}_{kl}}{\partial \os{\bar\g}{n}_{ij}}=\delta_{m,n} \mathbb{P}_{kl}^{ij}\qquad \forall m,n\geq 1 \, .
\end{equation}
where we introduced the following projector on the traceless part:
\begin{equation}
	\mathbb{P}_{ij}^{kl}=\delta_{(i}^k\delta_{j)}^l-\frac{1}{3}\os{\g}{0}_{ij}\os{\g}{0}^{kl} \, .
\end{equation}
This implies the $\os{\bar\gamma}{m}_{ij}$ are being varied only orthogonal to the direction of $\os{\gamma}{0}_{ij}$, i.e. 
\begin{equation}
	\os{\g}{0}^{ij}\frac{\delta}{\delta \os{\bar{\g}_{ij}}{m}}=0 \, .\label{subtlezero}
\end{equation}

Taking into account these subtleties, the variational principle for $\tilde S$ can then be related to the variational principle for $S$ via the standard procedure for changing variables. One finds the following relation between the two variational principles:
\begin{align}
	\frac{\delta \os{S}{n}}{\delta \os{\g}{0}_{ij}}&=\frac{\delta \os{\tilde S}{n}}{\delta \os{\g}{0}_{ij}}-\sum_{m=1}^n\left(\os{\bar\gamma}{m}^{ij}\frac{\delta \os{\tilde S}{n}}{\delta \os{\g}{m} }+\frac{\os{\gamma}{m}}{3}\left(\os{\gamma}{0}^{ij}\frac{\delta \os{\tilde S}{n}}{\delta \os{\g}{m}}+\frac{\delta \tilde S}{\delta \os{\bar \g}{m}_{ij}}\right)\right) \, , \label{partial1}\\
	\frac{\delta \os{S}{n}}{\delta \os{\gamma}{m}_{ij}}&=\os{\g}{0}^{ij}\frac{\delta \os{\tilde S}{n}}{\delta \os{\gamma}{m}}+\frac{\delta \os{\tilde S}{n}}{\delta \os{\bar\g}{m}_{ij}} \, .
\end{align}
Inversely one has
\begin{align}
	\frac{\delta \os{\tilde S}{n}}{\delta\os{\gamma}{0}_{ij}}&=\frac{\delta \os{S}{n}}{\delta \os{\g}{0}_{ij}}+\frac{1}{3}\sum_{m=1}^n\left(\os{\g}{m}^{ij}\os{\g}{0}_{kl}\frac{\delta \os{S}{n}}{\delta \os{\g}{m}_{kl}}+\os{\g}{0}^{mn}\os{\gamma}{n}_{mn}\mathbb{P}^{ij}_{kl} \frac{\delta \os{S}{n}}{\delta \os{\g}{m}_{kl}}\right)\, ,\\
	\frac{\delta \os{\tilde S}{n}}{\delta \os{\bar \g}{m}_{ij}}&=\mathbb{P}^{ij}_{kl}\frac{\delta \os{S}{n}}{\delta \os{\g}{m}_{kl}} \, ,\\
	\frac{\delta \os{\tilde S}{n}}{\delta \os{\g}{m}}&=\frac{1}{3}\os{\g}{0}_{kl}\frac{\delta \os{S}{n}}{\delta \os{\g}{m}_{kl}}\, .\label{partial2}
\end{align}

After we dealt with these technicalities, it has now become straightforward to prove \eqref{eqeoms}. Using the relations (\ref{partial1}-\ref{partial2}) one computes
\begin{align}
	\left.\os{\cale}{n}^{ij}\right|_{\os{\g}{m}=0}&=\left.\frac{\delta \os{S}{n}}{\delta \os{\gamma}{0}_{ij}}\right|_{\os{\g}{m}=0} \\
	&=\left.\frac{\delta \os{\tilde S}{n}}{\delta \os{\gamma}{0}_{ij}}\right|_{\os{\g}{m}=0}-\left.\sum_{p=1}^n\os{\bar\gamma}{p}^{ij}\frac{\delta \os{\tilde S}{n}}{\delta \os{\g}{p}}\right|_{\os{\g}{m}=0} \\
	&=\frac{\delta \bigg(\os{\tilde S}{n}\big|_{\os{\g}{m}=0}\bigg)}{\delta \os{\gamma}{0}_{ij}}-\frac{1}{3}\left.\sum_{p=1}^n\os{\bar\gamma}{p}^{ij}\os{\g}{0}_{kl}\frac{\delta \os{S}{n}}{\delta \os{\g}{p}_{ij}}\right|_{\os{\g}{m}=0}\label{keystep}\\
	&=\os{\bar \cale}{n}^{ij}-\frac{1}{3}\sum_{p=1}^n\os{\bar\gamma}{p}^{ij}\left.\os{\cale}{n-p}\right|_{\os{\g}{m}=0} \, .
\end{align}
Summarizing we find
\begin{equation}
	\os{\bar \cale}{n}^{ij}=\left.\left(\os{\cale}{n}^{ij}+\frac{1}{3}\sum_{p=1}^n\os{\bar\gamma}{p}^{ij}\os{\cale}{n-p}\right)\right|_{\os{\g}{m}=0}\label{traceandtracelesseom}  \, .
\end{equation}
This relation, together with (\ref{equiv1}, \ref{equiv2}), then immediately implies \eqref{eqeoms}.

\subsubsection{Summary}
The upshot is that the Euler-Lagrange equations $\{\os{\bar{\cale}}{n}_\psi=0,\os{\bar{\cale}}{n}^i=0, \os{\bar{\cale}}{n}^{ij}=0\}$ obtained by varying a particular, simpler, gauge-fixed action $\os{\bar{S}}{n}$ are equivalent to the expanded equations of motion $\{\os{{\cale}}{n}_\psi=0,\os{{\cale}}{n}^i=0, \os{{\cale}}{n}^{ij}=0\}$ in traceless gauge. We now summarize the precise definition of this alternative action and the related variational principle.

The gauge fixed actions $\os{\bar S}{n}$ are
\begin{eqnarray}
	\os{\bar S}{n}[\os{\psi}{0},\os{C}{0}_i,\os{\gamma}{0}_{ij};\os{\psi}{m},\os{C}{m}_i,\os{\bar\gamma}{m}_{ij}]&=&\left.\os{S}{n}[\os{\psi}{0},\os{C}{0}_i,\os{\gamma}{0}_{ij};\os{\psi}{m},\os{C}{m}_i,\os{\bar\gamma}{m}_{ij}]\right|_{\os{\g}{n}=0}\\&=&\os{\tilde S}{n}[\os{\psi}{0},\os{C}{0}_i,\os{\gamma}{0}_{ij};\os{\psi}{m},\os{C}{m}_i,\os{\bar\gamma}{m}_{ij},0] \, .
\end{eqnarray}
In practice the $\os{\bar S}{n}$ can be obtained by expanding the KS action (\ref{totlag}, \ref{L012}), replacing all $\os{\g}{m}_{ij}$ by $\os{\bar\g}{m}_{ij}$ and using that $\os{\gamma}{0}^{ij}\os{\bar\g}{m}_{ij}=0$.

The associated Euler-Lagrange equations are then defined as
\begin{equation}
	\os{\bar\cale}{n}_\psi=\frac{\delta \os{\bar S}{n}}{\delta \os{\psi}{0}}\, ,\qquad \os{\bar\cale}{n}^i=\frac{\delta \os{\bar S}{n}}{\delta \os{C}{0}_i} \, ,\qquad \os{\bar\cale}{n}^{ij}=\frac{\delta \os{\bar S}{n}}{\delta \os{\g}{0}_{ij}}+\frac{1}{3}\sum_{m=1}^n \frac{\delta \os{\bar S}{n}}{\delta \os{\bar\g}{m}_{kl}}\os{\gamma}{0}_{kl}\os{\bar\gamma}{m}^{ij}  \, .
\end{equation}
We might need to clarify the presence of the last terms in $\os{\bar\cale}{n}^{ij}$. From our argument above, see in particular $\eqref{keystep}$, one see that the variational principle used is that associated to $\os{\tilde S}{n}$. As pointed out in \eqref{subtle}, in that variational principle $\os{\bar\gamma}{m}_{ij}$ and $\os{\gamma}{0}_{ij}$ are not fully independent, i.e. under variation of $\os{\gamma}{0}_{ij}$ one has
\begin{equation}
	\delta \os{\bar\gamma}{m}_{ij}=\frac{1}{3}\os{\g}{0}_{ij}\os{\bar\gamma}{m}^{kl}\delta \os{\g}{0}_{kl}\, .\label{subtledelta}
\end{equation}
It is this effect that is responsible for the extra terms in $\os{\bar\cale}{n}^{ij}$. There is another point of view, which based on \eqref{subtlezero} implies that variations of $\os{\bar\gamma}{m}_{ij}$ are such that the extra terms vanish. At first thought these two point of views might seem contradictory, but they are actually perfectly equivalent if interpreted correctly. 

Let us consider an example of varying $\os{\g}{0}_{ij}$ from the first point of view:
\begin{align}
	\delta (\os{\bar\gamma}{n}_{ij} F^{ij})&=\delta \os{\bar\gamma}{n}_{ij} F^{ij}+ \os{\bar\gamma}{n}_{ij}\delta F^{ij}\nonumber\\
	&=\frac{1}{3}F\os{\bar\gamma}{m}^{kl}\delta \os{\g}{0}_{kl}+ \os{\bar\gamma}{n}_{ij}\delta (\bar F^{ij}+\frac{1}{3}\os{\g}{0}^{ij}F)\\
	&=\os{\bar\gamma}{n}_{ij}\delta \bar F^{ij} \, .\nonumber
\end{align}
We see that the extra terms, the $F^{ij}\delta \os{\bar\gamma}{n}_{ij}$ originating from \eqref{subtledelta}, seem to play a key role in removing the trace of $F_{ij}$ in the final result. But looking at the same calculation differently, we could observe that since $\os{\bar\gamma}{n}_{ij}$ is traceless we could have from the beginning written
\begin{eqnarray}
	\delta (\os{\bar\gamma}{n}_{ij} F^{ij})&=&\delta (\os{\bar\gamma}{n}_{ij} \bar F^{ij})\nonumber\\
	&=&\delta \os{\bar\gamma}{n}_{ij} \bar F^{ij}+ \os{\bar\gamma}{n}_{ij}\delta \bar F^{ij}\\
	&=&\os{\bar\gamma}{n}_{ij}\delta \bar F^{ij}\nonumber
\end{eqnarray}
In this way of doing the calculation the extra terms, now $\delta \os{\bar\gamma}{n}_{ij} \bar F^{ij}$, actually vanish.

Of the two approaches the second might seem the most efficient, namely to project from the beginning everything contracted with the $\os{\bar\gamma}{n}_{ij}$ on its traceless part so that one can ignore the extra contribution from $\delta \os{\bar\gamma}{n}_{ij}$. On the other hand, the first approach has the advantage of being more straightforward. In particular when time derivatives are present the first approach seems more tractable. In that case one can use, by \eqref{subtledelta}, that
\begin{equation}
	\delta \os{\dot{\bar\gamma}}{m}_{ij}=\frac{1}{3}\os{\dot{\g}}{0}_{ij}\os{\bar\gamma}{m}^{kl}\delta \os{\g}{0}_{kl}-\frac{1}{3}\os{\g}{0}_{ij}\os{\dot{\bar\gamma}}{m}^{kl}\delta \os{\g}{0}_{kl}+\frac{1}{3}\os{\g}{0}_{ij}\os{\bar\gamma}{m}^{kl}\delta \os{\dot\g}{0}_{kl}\label{subtledeltadot}
\end{equation}
This illustrates that a number of further terms appear, but partial integration will only be needed for the last one. 

Taking into account (\ref{subtledelta}, \ref{subtledeltadot}) then gives a precise way to compute all the equations of motion $\os{\bar \cale}{n}^{ij}$ by varying $\os{\bar S}{n}$ with respect to $\delta \os{\g}{0}_{ij}$.

\section{Extra trace terms}\label{traceapp}
In the main text we have chosen to present a number of results in a gauge where we have put to zero the traces $\os{\g}{n}$ of the subleading tensor fields $\os{\gamma}{n}_{ij}$, $n\geq 1$, see section \ref{tracesec} and appendix \ref{techtraceapp} for details on how and why this is done. For completeness we provide in this appendix the additional terms that would appear in these results if they were not gauge-fixed. This could be useful in case one would like to study certain features in a gauge independent fashion, or in another gauge.

\subsection*{Additions to results in section \ref{linsec}}
The linear part of the Lagrangian gets the additional trace terms in the $1/c$ expansion:
\begin{equation}
\os{\call}{n}_{\mathrm{lin}}=\left.\os{\bar\call}{n}_{\mathrm{lin}}\right|_{\os{\bar \gamma}{n}\!{}_{ij}\rightarrow \os{\gamma}{n}\!{}_{ij}}+\frac{1}{2}\os{\g}{n}\os{\call}{0}\,,\label{lincor}
\end{equation}
where $\os{\call}{0}$ can be found in \eqref{LOlag}.

The 2nd order linear differential operators appearing in the linear part in the $1/c$ expansion get the additional trace terms:
\begin{align}
\mathbb{D}^2\os{\psi}{n}=\,&\left.\overline{\mathbb{D}}^2\os{\psi}{n}\right|_{\os{\bar \gamma}{n}\!{}_{ij}\rightarrow \os{\gamma}{n}\!{}_{ij}} + \frac{1}{2}\os{\g}{n}\os{E}{0}_\psi +\frac{1}{2}\partial^i \psi D_i\os{\g}{n}  \, ,\\
\mathbb{D}^2\os{C}{n}^i=\,&\left.\overline{\mathbb{D}}^2\os{C}{n}^i\right|_{\os{\bar \gamma}{n}\!{}_{kl}\rightarrow \os{\gamma}{n}\!{}_{kl}}+ \frac{1}{2}\os{\g}{n}\os{E}{0}^i + \frac{1}{2} C^{ji}D_j \os{\g}{n} \, ,\\
\mathbb{D}^2\os{\g}{n}^{ij}=\, &\left.\overline{\mathbb{D}}^2\os{{\bar \g}}{n}^{ij}\right|_{\os{\bar \gamma}{n}\!{}_{kl}\rightarrow \os{\gamma}{n}\!{}_{kl}}-\frac{1}{3}\os{\g}{n}^{ij}\gamma_{kl}\os{E}{0}^{kl} + \frac{1}{2}\os{\g}{n} \os{E}{0}^{ij} \\
&+ \frac{1}{2}(D^{(i}D^{j)} \os{\g}{n} -  \g^{ij} D_k D^k \os{\g}{n})\,, \nonumber
\end{align}
where
\begin{align}
\os{E}{0}^{ij} =& \,-(R^{ij}+\frac{e^{2\psi}}{2}C_k{}^iC^{kj}-\frac{1}{2}\partial^i\psi\partial^j\psi)\\
&+\frac{\gamma^{ij}}{2}(R+\frac{e^{2\psi}}{4}C_{kl}C^{kl}-\frac{1}{2}\partial_k\psi\partial^k\psi) \nonumber
\end{align}
are leading order equations of motion, and for this reason can be ignored on-shell.

\subsection*{Additions to results in section \ref{explicitsec}}
The extra contributions to the linear parts of the Lagrangian are already given at all order in \eqref{lincor}.  In addition to those one has that $\os{L}{1}_{1\mathrm{td}}=\os{\bar L}{1}_{1\mathrm{td}}$  and
\begin{align}
\os{L}{2}_{0\mathrm{td}}=&\left.\os{\bar L}{2}_{0\mathrm{td}}\right|_{\bar \alpha_{kl}\rightarrow \alpha_{kl}}\!\!\!\!+\frac{\a}{2}\left(\frac{\a}{4}(R-\frac{1}{2}\partial_i\psi\partial^i\psi+\frac{e^{2\psi}}{4}C_{ij}C^{ij})\right.\label{extrashuffle1}\\
&\left.- \a^{ij}(R_{ij}-\frac{1}{2}\partial_i\psi\partial_j\psi+\frac{e^{2\psi}}{2}C_i{}^lC_{jl})-\partial_i\psi \partial^i\chi+\frac{\chi}{2} e^{2\psi}C_{ij}C^{ij}+\frac{e^{2\psi}}{4}B_{ij}C^{ij}\right)\nonumber  \, ,\\
\os{L}{2}_{1\mathrm{td}}=&\left.\os{\bar L}{2}_{1\mathrm{td}}\right|_{\bar \alpha_{kl}\rightarrow \alpha_{kl}}\!\!\!\!+\frac{\a}{2}\left(\G^{ij\,kl} D_i C_j \dot{\gamma}_{kl}
+ C_i (\dot{\psi}\partial^i \psi  -e^{2\psi} \dot{C}_j C^{ij})  + 2 \dot{C}_i \partial^i\psi\right)\label{extrashuffle2}\\
&	- \frac{1}{2}C_i \dot{\gamma} D^i \a   \, , \nonumber \\
\os{L}{2}_{2\mathrm{td}} =&\left.\os{\bar L}{2}_{2\mathrm{td}}\right|_{\bar \alpha_{kl}\rightarrow \alpha_{kl}}\!\!\!\!   + \frac{\a}{2}e^{-2\psi}\left(\frac{1}{4}\Gamma^{kl\,mn} \dot \gamma_{kl}\dot \gamma_{mn}+ \dot \gamma \dot \psi -\frac{3}{2} \dot{\psi}^2\right) \, .
\end{align}

\subsection*{Additions to results in section \ref{evlinsec}}
The linear part of the Lagrangian gets the additional trace terms in the $1/c^2$ expansion:
\begin{equation}
\os{\call}{2n}_{\mathrm{lin}}^{\mathrm e}=\left.\os{\bar\call}{2n}_{\mathrm{lin}}^{\mathrm e}\right|_{\os{\bar \gamma}{n}\!{}_{ij}\rightarrow \os{\gamma}{2n}\!{}_{ij}}+\frac{1}{2}\os{\g}{2n}\os{\call^{\mathrm e}}{0}\, ,\label{lincor2}
\end{equation}
where $\os{\call}{0}^\mathrm{e}$ can be found in \eqref{LOevlag}.

The 2nd order linear differential operators appearing in the linear part in the $1/c^2$ expansion get the additional trace terms:
\begin{align}
\mathbb{D}^2\os{\psi}{2n}=\,&\left.\overline{\mathbb{D}}^2\os{\psi}{2n}\right|_{\os{\bar \gamma}{2n}\!{}_{ij}\rightarrow \os{\gamma}{2n}\!{}_{ij}} + \frac{1}{2}\os{\g}{2n}\os{E}{0}_\psi^{\mathrm e} +\frac{1}{2}\partial^i \psi D_i\os{\g}{2n}  \, ,\\
\mathbb{D}^2\os{\g}{2n}^{ij}=\,&\left.\overline{\mathbb{D}}^2\os{{\bar \g}}{2n}^{ij}\right|_{\os{\bar \gamma}{2n}\!{}_{kl}\rightarrow \os{\gamma}{2n}\!{}_{kl}}-\frac{1}{3}\os{\g}{2n}^{ij}\gamma_{kl}\os{E}{0}^{kl}_{\mathrm e} + \frac{1}{2}\os{\g}{2n} \os{E}{0}^{ij}_{\mathrm e} \\
&+ \frac{1}{2}(D^{(i}D^{j)} \os{\g}{2n} - \g^{ij} D_k D^k \os{\g}{2n})\,, \nonumber\\
 \cald_{\mathrm e}^2\os{B}{2n+2}^i&=\left.\overline{\cald}_{\mathrm e}^2\os{B}{2n+2}^i\right|_{\os{\bar \gamma}{2n}\!{}_{kl}\rightarrow \os{\gamma}{2n}\!{}_{kl}} \, ,
\end{align}
where
\begin{equation}
\os{E}{0}^{ij}_{\mathrm{e}}=-(R^{ij}-\frac{1}{2}\partial^i\psi\partial^j\psi)+\frac{\gamma^{ij}}{2}(R-\frac{1}{2}\partial_k\psi\partial^k\psi)
\end{equation}
are leading order equations of motion. 

\subsection*{Additions to results in section \ref{evexplsec}}
The extra contributions to the linear parts of the even Lagrangian are already given at all order in \eqref{lincor2}.  Outside the traceless gauge, the Lagrangians (\ref{L2evlag},
	\ref{evnnlo}) become
\begin{align}
\os{L}{2}^\mathrm{e}=&\left.\os{\bar L}{2}^\mathrm{e}\right|_{\bar \b_{kl}\rightarrow \b_{kl}}\!\!\!\! + \frac{\b}{2} \left(R-\frac{1}{2}\partial_i\psi\partial^i\psi\right)\, ,\label{superimportanttrace}\\
\os{L}{4}^\mathrm{e}=&\left.\os{\bar L}{4}^\mathrm{e}\right|_{\substack{\bar \b_{kl}\rightarrow \b_{kl}\\\bar \e_{kl}\rightarrow \e_{kl}}}\!\!\!\! + \frac{\e}{2} \left(R-\frac{1}{2}\partial_i\psi\partial^i\psi\right)  \\
& +\frac{\b}{2}\left(\frac{\b}{4}(R-\frac{1}{2}\partial_i\psi\partial^i\psi ) -\b^{ij}(R_{ij}-\frac{1}{2}\partial_i\psi\partial_j\psi)-\partial_i\psi \partial^i\phi+\frac{e^{2\psi}}{4}B_{ij}B^{ij}\right. \nonumber\\
 & \left.+\G^{ij\,kl} D_i B_j \dot \g_{kl}+ B_i \dot{\psi}\partial^i \psi   + 2 \dot{B}_i \partial^i\psi + \frac{e^{-2\psi}}{4}\left(\Gamma^{ij\,kl} \dot \gamma_{ij}\dot \gamma_{kl}+4 \dot \gamma \dot \psi-6 \dot{\psi}^2\right)\right)\nonumber\\
&- \frac{1}{2}B_i \dot{\gamma} D^i \b  \, .\nonumber
\end{align}
\section{Equations of motion}\label{eomapp}

Here we present the equations of motion of the (un-expanded) KS action. As in (\ref{totlag}) the Lagrangian can be split as
\begin{equation}
\call=\call_0+c^{-1}\call_1+c^{-2}\call_2\qquad \call_a=\sqrt{\g}L_a \, ,
\end{equation}
with the $L_a$ given in (\ref{L012}). Similarly we can split the equations of motion (in the absence of matter) as
\begin{equation}
E_0=-c^{-1}E_1-c^{-2}E_2\qquad E_a=\frac{1}{\sqrt{\g}}\frac{\delta \call_a}{\delta \Phi} \, .
\end{equation}
We present the different parts of the equations of motion obtained from variation by each of the fields $\Phi=(\psi,C_i,\gamma_{ij})$ below.

\subsection*{Variation by $\psi$}
Variation of the KS Lagrangian with $\psi$ yields
\begin{align}
E_0^\psi =& \,\,D_i\partial^i \psi +\frac{e^{2\psi} }{2} C_{ij} C^{ij} \, , \\
- E_1^\psi =& \,\,D_i (C^i \dot{\psi}) + \dot{C}^i \partial_i \psi + 2 D_i \dot{C}^i + C^i \Big( \partial_i \dot{\psi} + \frac{\dot{\gamma}}{2} \partial_i \psi - \dot{\gamma}_{ij} \partial^j \psi \Big) \\
&+ 2 e^{2\psi} C_{ij} C^i \dot{C}^j \, , \nonumber \\
- E_2^\psi =&   - 2 \dot{\psi} C^i \dot{C}_i - 2 C^i\ddot{C}_i - 2 \dot{C}^i\dot{C}_i - C^iC_i\ddot{\psi} \\
&- \Big(  C_i \frac{\dot{\gamma}}{2} - C^j\dot{\gamma}_{ij} \Big) (\dot{\psi}C^i + 2 \dot{C}^i)  + e^{2\psi} \Gamma^{ij\,kl} C_i C_j \dot{C}_k \dot{C}_l\nonumber \\
&- \frac{1}{2} e^{-2\psi} \Big(  \dot{\gamma}^{ij} \dot{\gamma}_{ij} + 3 \dot{\gamma}\dot{\psi} -6 \dot{\psi}^2 +6 \ddot{\psi} -2 \gamma^{ij} \ddot{\gamma}_{ij}   \Big)\nonumber \, .
\end{align}
\subsection*{Variation by $C_i$}
Next, one finds the equation of motion for the gauge field $C_i$ to be
\begin{align}
E_0^i =&\,\, D_j (e^{2\psi}  C^{ij}) \, ,  \\
- E_1^i =& \,\, \Gamma^{ij\,kl} D_j \dot{\gamma}_{kl} - \dot{\psi} \partial^i \psi + \dot{\gamma} \partial^i\psi + 2 \partial^i \dot{\psi} - 2 \dot{\gamma}^{ij} \partial_j \psi + 2 D_j (e^{2\psi} C^{[i} \dot{C}^{j]})\\
&+ e^{2\psi} \left(  \Big( 2\dot{\psi} C_j +2 \dot{C}_j + \frac{1}{2} \dot{\gamma}C_j - C^k\dot{\gamma}_{jk}  \Big) C^{ij} + C_j \dot{C}^{ij}  + C_j \dot{\gamma}_k{}^i C^{jk}  \right)\, ,\nonumber\\
- E_2^i =& -\Gamma^{ij\,kl} C_j \ddot{\gamma}_{kl} - \frac{1}{2} (C^i \dot{\gamma}^{jk}\dot{\gamma}_{jk} + \dot{\gamma}^{ij} C_j \dot{\gamma} )+C^k\dot{\gamma}^{ij}\dot{\gamma}_{jk} \\
&+\dot{\psi}^2 C^i - \dot{\gamma}C^i \dot{\psi} - 2 C^i \ddot{\psi} +2 C_j \dot{\psi} \dot{\gamma}^{ij} \nonumber \\
&-2 e^{2\psi} \left(  \Big( 2 C_j \dot{\psi} + 2 \dot{C}_j  - C^k \dot{\gamma}_{jk} + \frac{1}{2}C_j \dot{\gamma}  \Big) C^{[i} \dot{C}^{j]}      + C_j \Big( C^{[i} \ddot{C}^{j]} + \dot{\gamma}^i{}_{k} C^{[j}\dot{C}^{k]}  \Big)    \right)\, .  \nonumber
\end{align}

\subsection*{Variation by $\g_{ij}$}
It is a bit cumbersome but straightforward to obtain the contributions:
\begin{align}
E_0^{ij} =& \,\,\frac{\gamma^{ij}}{2} \Big(R - \frac{1}{2} \partial_k \psi \partial^k\psi + \frac{e^{2\psi}}{4}C_{kl}C^{kl}\Big)- \Big(R^{ij} - \frac{1}{2} \partial^{i}\psi \partial^{j} \psi + \frac{e^{2\psi}}{2} C^{i}{}_{k}C^{jk} \Big) ,\\ 
- E_1^{ij} =& \,\,\frac{\gamma^{ij}}{2} \Gamma^{kl\,mn} (\dot{\gamma}_{mn} D_k C_l + 2 C_k D_l \dot{\gamma}_{mn}) + C^kD_k \dot{\gamma}^{ij} - C^{(i}D_k \dot{\gamma}^{j)k} \\
& - C_k D^{(i} \dot{\gamma}^{j)}{}_{k} +C^{(i} D^{j)} \dot{\gamma} + \frac{\dot{\gamma}^{ij}}{2}D_k C^k - \dot{\gamma}^{k(i} D_k C^{j)}  +\frac{\dot{\gamma}}{2} D^{(i} C^{j)}    \nonumber \\
&+  \frac{e^{2\psi}}{2} \Big(  C_{kl} (\frac{\gamma^{ij}}{2} C^k \dot{C}^l - 2 \gamma^{ki} C^{[j} \dot{C}^{l]} )   + C_{[k} \dot{C}_{l]} (\frac{\gamma^{ij}}{2} C^{kl} - 2 \gamma^{k(i} C^{j)l})    \Big)     \nonumber \\  
& + \Gamma^{ij\,kl} D_k\dot{C}_l - \gamma^{ij}\dot{C}^k \partial_k \psi + 2\dot{C}^{(i} \partial^{j)}\psi- \dot{\psi}\Big(   \frac{{\gamma}^{ij}}{2} C^k\partial_k\psi - C^{(i}\partial^{j)} \psi \Big) \, , \nonumber\\
-E_2^{ij} =& \,\, C_k C^{(i}\ddot{\gamma}^{j)k} -\frac{1}{2} C^k C_k \ddot{\gamma}^{ij} - \frac{1}{2} C^{i} C^{j}\ddot{\gamma} + C_k \dot{C}^{(i} \dot{\gamma}^{j)k} - \frac{1}{2} C^k \dot{C}_k \dot{\gamma}^{ij} \\
& + \frac{1}{4} C^{(i} C^{j)} \dot{\gamma}^{kl}\dot{\gamma}_{kl} - \frac{1}{2} \dot{C}^{(i} C^{j)} \dot{\gamma}  + \frac{1}{2} C_k C_l \dot{\gamma}^{kl}\dot{\gamma}^{ij} - \frac{1}{4} C_k C^k \dot{\gamma}\dot{\gamma}^{ij} - C^k C^{(i}  \dot{\gamma}^{j)l} \dot{\gamma}_{kl}\nonumber \\
&+ \frac{1}{2} C_k  C^{(i} \dot{\gamma}^{j)k}\dot\gamma - \frac{1}{2} C_k C_l \dot{\gamma}^{k(i} \dot{\gamma}^{j)l} + \frac{1}{2} C_k C^k \dot{\gamma}^{l(i} \dot{\gamma}^{j)}{}_l\nonumber\\
& - \Gamma^{(ij)\,kl} \Big( C_k \ddot{C}_l + \frac{1}{2} \dot{C}^m (C_m \dot{\gamma}_{kl} - 2 C_k \dot{\gamma}_{lm}) + \dot{C}_k \dot{C}_l   \Big) +2 \dot{\psi} (\frac{\gamma^{ij}}{2} \dot{C}_k C^k - \dot{C}^{(i} C^{j)}) \nonumber \\
&+ \frac{1}{4} ( \gamma^{ij} C_k C^k - 2C^{i} C^{j} ) \dot{\psi}^2 - e^{2\psi} C_{[k} \dot{C}_{l]} (\frac{\gamma^{ij}}{2} C^k \dot{C}^l -2 \gamma^{ik} C^{[j} \dot{C}^{l]}  ) \nonumber \\
&- \frac{e^{-2\psi}}{2}  \gamma^{ij} \Big(\frac{\dot{\gamma}^2}{4}   - \frac{3}{4} \dot{\gamma}_{kl} \dot{\gamma}^{kl}   -2 \dot{\psi} \dot\gamma + \frac{5}{2} \dot{\psi}^2\Big)\nonumber \\
&-\frac{e^{-2\psi}}{2}  \Big( \dot{\gamma}^{k(i} \dot{\gamma}^{j)}{}_k + (2\dot{\psi}- \frac{\dot{\gamma}}{2})\dot{\gamma}^{ij}   -2 \gamma^{ij} \ddot{\psi} - \Gamma^{ij\,kl}\ddot{\gamma}_{kl}\Big)\nonumber \\
&- \frac{\gamma^{ij}}{2}\Big(  \Gamma^{kl\,mn} (C_k \dot{C}_l \dot{\gamma}_{mn} + C_k C_l \ddot{\gamma}_{mn}) - \frac{1}{4} C_kC^k \dot{\gamma}^2 + C_k C_l \dot{\gamma}\dot{\gamma}^{kl}  \Big) \nonumber\\
& - \frac{\g^{ij}}{2} \Big(\frac{3}{4} C_k C^k \dot{\gamma}^{lm}\dot{\gamma}_{lm}    - \frac{3}{2} C^k C^l \dot{\gamma}_{km}\dot{\gamma}^{m}{}_{l}  \Big) \, . \nonumber
\end{align}

\bibliographystyle{utphys}
\bibliography{31dual}

\end{document}